\documentclass[aps,pra,onecolumn,showpacs,amsmath,amssymb,floatfix,superscriptaddress,notitlepage, longbibliography]{revtex4-2}
\usepackage[T1]{fontenc} % codifica dei font
\usepackage[utf8]{inputenc}
\usepackage[english]{babel}
\usepackage{xcolor}

\usepackage{graphicx}
\usepackage{epsfig}
\usepackage{subfigure}
\usepackage{dcolumn}
\usepackage{bm}
\usepackage{graphicx,bbm,psfrag}
\usepackage{amsmath} % Required for inserting images
\usepackage{amsfonts}
\usepackage{yfonts}
\usepackage[protrusion=true,expansion=true]{microtype}
\usepackage{times}
\usepackage{amssymb}
\usepackage{hyperref}
\usepackage{physics}

\newcommand{\bc}{\begin{center}}
\newcommand{\ec}{\end{center}}
\def\ba#1{\begin{array}{#1}\displaystyle}
\newcommand{\ea}{\end{array}}
\newcommand{\z}{\\ \displaystyle}
\newcommand{\beq}{\begin{equation}}
\newcommand{\eeq}{\end{equation}}
\newcommand{\beqa}{\begin{eqnarray}}
\newcommand{\eeqa}{\end{eqnarray}}

\newcommand{\bi}{\begin{itemize}}
\newcommand{\ei}{\end{itemize}}

\newcommand{\R}{{\mathbb{R}}}

\newcommand{\ep}{\epsilon}
\newcommand{\varep}{\varepsilon}
\newcommand{\q}{\theta}

\newcommand{\be}{\begin{equation}}
\newcommand{\ee}{\end{equation}}

\date{\today}

\begin{document}

\author{D\'avid X. Horv\'ath}
\author{Bruno Bertini}
\affiliation{School of Physics and Astronomy, University of Birmingham, Edgbaston, Birmingham, B15 2TT, UK}

\title{Large-scale dynamics of integrable quenches}

\begin{abstract}
The Ballistic Macroscopic Fluctuation Theory (BMFT) is a path-integral based approach to treat large-scale correlations in interacting integrable systems which, so far, has only been applied to quasi-equilibrium settings. In this paper we extend this approach to integrable quench problems and, to illustrate it, we compute the time evolution of the full-counting-statistics (FCS) of a conserved charge. We identify the relevant path integral, solve it via saddle-point, and obtain a set of partial differential equations describing the dynamics of the FCS. We solve these equations explicitly for free fermionic systems and for the interacting cellular automaton Rule 54. In both cases, we recover the known exact results. For Rule 54, our construction moreover completes the solution across the entire ballistic window.
\end{abstract}

\maketitle

\section{Introduction}
\label{sec:intro}

Whenever a quantum many-body system with short range interactions is let to evolve for long enough times, its spatially local subsystems almost entirely forget the initial configuration. Namely, their state eventually takes a Gibbs-like form determined by the local conservation laws of the system, where the only dependence on the initial condition is in the chemical potentials (Lagrange multipliers)~\cite{polkovnikov2011colloquium, eisert2015quantum, calabrese2016introduction, serbyn2021quantum, bastianello2022introduction, abanin2019colloquium}. The latter are static in homogeneous situations and follow (classical) hydrodynamic flow equations in the presence of spatial inhomogeneity~\cite{castroalvaredo2016emergent,bertini2016transport,doyon2020lecture,alba2021generalized,essler2023a}. This can be considered as an instance of universality, where the behaviour of the system is described by an emergent macroscopic theory and the only information about the initial configuration (``the microscopics'') is contained in a set of coefficients. 

A fascinating question is whether similar forms of universality can also emerge at finite times, i.e., when local subsystems are far from any equilibrium state and the physics is inherently richer. Answering this question is substantially harder than its equilibrium counterpart because of the lack of general methods to access the finite time regime in the presence of non-trivial interactions. Until very recently, this was the case also for highly structured systems like integrable models: although integrability gave access to their thermodynamics~\cite{takahashi2005thermodynamics, korepin1997quantum} and stationary properties~\cite{caux2016the, essler2016quench} their finite time dynamics remained completely out of reach. Recent results, however, have started to change this picture. On the one hand, researchers have discovered examples of integrable models with exactly solvable dynamics despite the non-trivial interactions~\cite{pozsgay2014quantum,pozsgay2016real,zadnik2021foldedI,zadnik2021foldedII, giudice2022temporal, klobas2021exact,klobas2021exact2,klobas2021entanglement,klobas2024non, yang2026solving, klobas2026exact, travaglino2026dynamical}, a chief example being the quantum cellular automaton Rule 54~\cite{bobenko1993two, buca2021rule54}. On the other hand, they have developed general approaches to treat large-scale dynamics in integrable models based upon exploiting a duality between space and time~\cite{bertini2022growth, bertini2023nonequilibrium, bertini2024dynamics, travaglino2026space} or on extending the hydrodynamic description beyond the regime of asymptotically large times~\cite{horvath2025counting,horvath2025a,del_Vecchio_del_Vecchio_2024,li2026, Del_Vecchio_Del_Vecchio_2022}. Due to the scarcity of independent checks, however, it is very important to cross check and benchmark the predictions of these approaches to validate the assumptions they rely upon.

In this spirit, here we present a description of non-equilibrium dynamics by means of the so called Ballistic Macroscopic Fluctuation Theory (BMFT)~\cite{doyon2023bmft}. The latter is a hydrodynamics-based approach developing a path-integral characterisation of large-scale correlations that, so far, has only been applied to equilibrium or quasi-equilibrium settings \cite{doyon2023bmft, doyon2020fluctuations,perfetto2021euler, yoshimura2025anomalous, kethepalli2025bmft,Hubner_2025,  Yoshimura_2026, Yoshimura_2026B}. We extend this description to the genuinely out-of-equilibrium setting of quantum quenches and apply it to the ballistic-scale full counting statistics (FCS) of a conserved charge~\cite{eisler2013full,perfetto2020dynamics,Bertini_2023_tFCS,bertini2024dynamics,horvath2024fcs,Senese_2024,horvath2025counting}. Although conserved in the full system, the charge exhibits nontrivial fluctuations and dynamics when restricted to a large subsystem after the quench. In particular,  we define
\beq
f(\lambda, \mu, X,T)=X^{-1}\ln\langle\Psi|e^{\lambda \int_0^X \dd x\, q(x,T)}|\Psi\rangle\,,
\eeq
where $\lambda$ is the counting field, $\mu$ is a set of coordinates that describe the state, $X$ denotes the length of the subsystem, and  $T$ is the time elapsed after the quantum quench. At a later stage in this work $X$ is eventually sent to infinity, therefore to probe the ballistic scale fluctuations, $T$ is set to be proportional to $X$ as $T=\zeta X$, and the eventual FCS only depends on this ratio $\zeta$.
Our main result is an expression of its evolution in terms of a set of saddle point equations (classical PDEs). We solve these equations exactly in the cases of free fermions (non-interacting) and Rule 54 (interacting) and, in both cases, we recover known exact results. For Rule 54 we also extend the validity of the exact result to the entire ballistic regime. 

The rest of the paper is laid out as follows. In Sec.~\ref{sec:setting} we discuss our setting, briefly reviewing the main ideas behind the thermodynamic description of integrable models and discuss the initial states considered. In Sec.~\ref{BMFTIntro} we present the main ideas of BMFT and of its extension to the non-equilibrium setting. In Sec.~\ref{sec:time_dep_FCS} we present a BMFT characterisation of the FCS and discuss the special cases of free fermions and Rule 54, where the equations can be solved exactly. Finally, Sec.~\ref{sec:conclusions} contains our conclusions.

\section{Setting}
\label{sec:setting}

\subsection{Thermodynamic Description of Bethe-Ansatz Integrable Models}
\label{TBA}

In this work we focus on Bethe ansatz integrable systems~\cite{korepin1997quantum}: a class of quantum many-body systems, generically interacting, but where the scattering is elastic and can be factorised into sequences of two-particle processes. This implies that integrable systems are characterised by an extensive number of conserved charges with local density, denoted by $\{Q_k\}$, and that their spectrum can be described in terms of stable quasiparticle excitations~\cite{korepin1997quantum}. The $j$-th quasiparticle carries energy $\ep(\vartheta_j)$ and momentum $p(\vartheta_j)$ that are conveniently parameterised in terms of a rapidity $\vartheta_j$. The eigenstates $\ket{\{\vartheta_j\}}$ of these systems can be written as scattering states (superpositions of plain waves) with a fixed set of quasiparticles $\{\vartheta_j\}$. Although in the presence of interactions the structure of eigenstates can be very complicated, expectation values of conserved quantities are expressed as simple sums over $\{\vartheta_j\}$. For instance, the momentum and energy associated with a given eigenstate are
\be
P\left[\{\vartheta_j\}\right]=\sum_j p(\vartheta_j)\,,\qquad E\left[\{\vartheta_j\}\right]=\sum_j \varepsilon(\vartheta_j)\,.
\ee

As it is customary in quantum mechanics, whenever the system is confined to a finite volume the values of the momenta, and therefore of the rapidities, are quantised. Due to the interactions, however, the quantisation conditions couple all rapidities together. For an $N$-quasiparticle state the quantisation conditions, known as Bethe equations, take the form~\cite{korepin1997quantum}
\be\label{eq:bethe_eq}
e^{i L p(\vartheta_r)}=\mathcal{S}_r(\vartheta_1\,\ldots, \vartheta_N)\,,\qquad r=1\,\ldots N,
\ee
where $\mathcal{S}_r$ depend on the specific model considered. In general, the solutions to these equations need not to be real, and can take arbitrary values in the complex plane. For large volumes, however, the rapidities $\{\vartheta_j\}$ arrange themselves in regular patterns in the complex plane where the rapidities stay at a fixed distance from one another~\cite{takahashi2005thermodynamics}. The latter can be specified by a real rapidity and are interpreted as bound states formed by the elementary quasiparticles. One can then obtain quantisation conditions only involving real rapidities, but allowing for different species of quasiparticles (original ones and bound states). These new quantisation conditions are known as Bethe--Takahashi equations~\cite{takahashi2005thermodynamics}.

In the thermodynamic limit, the allowed values for the real rapidities become dense and one can describe them in terms of sets of distributions $\rho_{t,n}(\vartheta)$, where $n$ labels different species of quasiparticles (the specific set is model dependent). Due to the non-trivial coupling implemented by the quantisation conditions, these distributions depend on the portions $\rho_n(\vartheta)$ of rapidities that are occupied. The latter provide a ``macroscopic'' description of the system's eigenstates, i.e., a given set of $\rho_n(\vartheta)$s describes exponentially many microscopic eigenstates whose rapidities approach those distributions in the thermodynamic limit. Importantly, in the thermodynamic limit expectation values of local observables in any such eigenstate are fully specified by the rapidity distributions~\cite{korepin1997quantum}. The precise relation between $\rho_n(\vartheta)$ and $\rho_{t,n}(\vartheta)$ is found to be~\cite{takahashi2005thermodynamics, korepin1997quantum}
\be
\label{eq:rho-rhoTot_ext}
\rho_{t,n}(\vartheta) =\frac{p'(\vartheta)}{2\pi}+ \sum_{m}\int {\rm d}\varphi\, {T}_{nm}(\vartheta,\varphi) \rho_m(\varphi),
\ee
where $(\cdot)'$ denotes a derivative with respect to $\vartheta$ and we introduced the \emph{scattering kernel} 
\be
{T}_{nm}(\vartheta,\varphi) = -\frac{i}{2\pi }\frac{{\rm d}}{{\rm d}\vartheta} S_{nm}(\vartheta,\varphi), 
\ee
with $S_{nm}(\vartheta,\varphi)$ denoting the S-matrix describing the scattering between $(n,\vartheta)$ and $(m,\varphi)$. In the following we will use a more compact notation via introducing $\theta$ as a multi-index for $(n,\vartheta)$ and where Einstein's convention is interpreted as
\beq
a^\q \cdot b_\q=\sum_n \int \dd \vartheta\, a_n(\vartheta)b_n(\vartheta)\,,
\eeq
where upper-index quantities are regarded as row vectors and lower index quantities as column vectors. In this new notation we represent Eq.~\eqref{eq:rho-rhoTot_ext} as 
\beq
\rho_{\text{t},\q}=\frac{p'_\q}{2\pi}+\mathtt{T}_{\q}^{\,\,\phi} \cdot \rho_\phi\,, 
\label{rho-rhoTot}
\eeq
where the first, lower index of $\mathtt{T}$ is a row index, and the second upper index is a column index. Whereas distinguishing row and column vectors is convenient for subsequent manipulations,  the entries of vectors with upper and lower indices are the same since the metric tensor is the identity here.

One can find the rapidity distributions characterising distinguished eigenstates. For example, the root distribution providing a microcanonical description of the generalised Gibbs ensemble (GGE)
\be
\rho_{\rm GGE} \propto \exp[ -\mu^k Q_k], 
\label{eq:GGE}
\ee
is given by~\cite{takahashi2005thermodynamics, korepin1997quantum} 
\beq
\ep^{\q}=\mu^{\q}-\ln\left[1+e^{-\ep^{\phi}}\right] \cdot \mathtt{T}
_{\phi}^{\,\,\,\q} \,. 
%\quad \text{or} \quad \ep_{\q}=\mu_{\q}- \mathtt{T}_{\q}^{\,\,\,\phi}\cdot \ln\left[1+e^{-\ep_{\phi}}\right] \,,
\label{TBAEq}
\eeq
%depending on whether $\mu$ and $\epsilon$ are regarded as column or row vectors, but we shall typically use the former choice; and we used that $\mathtt{T}$ is symmetric. 
Here the driving term is expressed as 
\be
\mu^\q=\mu_n({\vartheta})=\sum_{k} \mu_k q_{k,n}(\vartheta), 
\label{RapidityDepMu}
\ee
where $q_{k,n}(\vartheta)$ is the contribution of the quasiparticle $(n,\vartheta)$ to the charge $Q_k$. Instead, $\ep^\q=\ep_n({\vartheta})$ is the \emph{pseudo-energy} defined in terms of the rapidity distributions as  
\be
{\ep^\q} =\log \frac{1-n^\q}{n^\q}
\label{PseudoE}
\ee
where we introduced the \emph{filling function} 
\beq
n_\q=\frac{\rho_\q}{\rho_{\text{t},\q}}\,.
\label{FillingFunction}
\eeq
Using the pseudo-energy, one can express the free energy density $\textgoth{f}$ of the GGE in Eq.~\eqref{eq:GGE} as
\beq
\textgoth{f}=-\ln\left[1+e^{-\ep^{\q}}\right]\cdot\frac{1_\q}{2\pi}=-\textgoth{f}^\q\cdot\frac{1_\q}{2\pi}\,,
\label{FreeEenergyDens}
\eeq
where we defined the rapidity-resolved free energy density (note the minus sign).
We also introduce the $\mathtt{R}$ matrix
\beq
\mathtt{R}=(1- n\mathtt{T})\,,\qquad
\mathtt{R}^T=(1-\mathtt{T}n)\,,
\label{RMatrix}
\eeq
where $^T$ denotes transposition and we used that $\mathtt{T}$ is symmetric.
Via $\mathtt{R}$ we define the dressing operations for column and row vectors, respectively
\beq
f_{\text{dr},\q}=\left(\mathtt{R}^{-T}\right)_{\q}^{\,\,\,\,\phi} \cdot f_\phi\,,\qquad
f^{\text{dr},\q}= f^\phi \cdot \left(\mathtt{R}^{-1}\right)_{\phi}^{\,\,\,\,\q}\,.
\label{DressingOps}
\eeq
The effective velocity and the flux Jacobian, which govern the Euler-scale hydrodynamics, are
\beq
v^{\text{eff}}_\q=\frac{(E')^{\text{dr}}_\q}{(p')^{\text{dr}}_\q}\,,\qquad
\mathtt{A}_\q^{\,\,\phi}=\frac{\delta (v^{\text{eff}}_\q \rho_\q)}{\delta \rho_\phi}\,,
\label{FluxJacobian}
\eeq
and are linked by the relation
\beq
\mathtt{R}\,\mathtt{A}\,\mathtt{R}^{-1}=\text{diag}\{v^\text{eff}_\q\}\,.
\label{RARInv}
\eeq
Let us conclude this brief summary by specialising the above treatment to two systems that will be considered in detail in the upcoming sections.

\subsubsection{Free Fermions}

In free fermionic systems the quasiparticles do not interact and, accordingly, the scattering matrix is trivial: $S_{\q,\phi}=1$. This implies that the scattering kernel vanishes, $\mathtt{T}=0$, and Eq.~\eqref{rho-rhoTot} reduces to
\beq
\rho_{\text{t},\q}=\frac{p'_\q}{2\pi}\,,
\eeq
i.e., the total density of states coincides with the bare one and is independent of the occupation. Consequently, the $\mathtt{R}$ matrix is the identity and all dressing operations are trivial: $f^{\text{dr},\q}=f^\q$ for any function $f$.

In this case the rapidity can be taken to coincide with the momentum, i.e.\ without loss of generality we can take $p'_\q=1$, but we still denote it by $\q$. The dispersion relation is then indicated by $\varep_\q$. In the following we will leave this function unspecified: we only assume it to be an even function of the rapidity. The effective velocity then equals the bare group velocity, i.e. 
\beq
v^{\text{eff}}_\q=\varep'_\q\,,
\eeq
and the flux Jacobian~\eqref{FluxJacobian} is diagonal: $\mathtt{A}=\text{diag}\{\varep'_\q\}$. Since $\mathtt{T}=0$, the TBA equation~\eqref{TBAEq} becomes simply $\ep_\q=\mu_\q$ and the filling function~\eqref{FillingFunction} reads
\beq
n_\q=\frac{1}{1+e^{\mu_\q}}\,.
\eeq
Finally, the free energy density~\eqref{FreeEenergyDens} takes the familiar form
\beq
\textgoth{f}=-\int\frac{\dd \q}{2\pi}\ln\left(1+e^{-\mu_\q}\right)\,.
\eeq

\subsubsection{Rule 54}

Rule~54 is a deterministic, time-reversible cellular automaton defined on a one-dimensional lattice with binary site variables $s_x\in\{0,1\}$~\cite{bobenko1993two}. The dynamics is local and proceeds in discrete time steps (which is equivalent to a Floquet evolution): at each half time-step the state of every other site $x$ is updated according to
\beq
s_x\!\bigl(t{+}\tfrac{1}{2}\bigr)
=s_x(t)\oplus s_{x-1}(t)\oplus s_{x+1}(t)
  \oplus s_{x-1}(t)\,s_{x+1}(t) \,,
\eeq
where $\oplus$ denotes addition modulo two;
the middle bit is flipped whenever at least one of its two neighbours is in state~$1$.
Despite the simplicity of this rule, the model supports stable quasiparticles---left- and right-moving solitons that propagate ballistically and interact through a simple position shift upon scattering~\cite{klobas2019time,buca2021rule54}. 
This coexistence of ballistic soliton propagation with a nontrivial, factorised scattering makes Rule~54 one of the simplest \emph{interacting} integrable models, and an ideal testbed for exact methods in nonequilibrium statistical mechanics~\cite{buca2021rule54,gopalakrishnan2018operator,friedman2019integrable}.

At the hydrodynamic level, the two soliton species---equivalently viewed as two discrete rapidity modes $\nu=\pm1$ of a single particle type---give rise to a two-component TBA.
The scattering kernel is a $2\times 2$ matrix,
\beq
\mathtt{T}=\begin{pmatrix}
-1&1\\
1&-1\\
\end{pmatrix}\,,
\eeq
from which all remaining TBA quantities follow. The quasiparticle densities are denoted by $\rho_+$ and $\rho_-$ and Eq.~\eqref{rho-rhoTot} takes the form 
\beq
\rho_{\text{t}, \nu}=1-\rho_\nu+\rho_{-\nu}\,,
\label{rhoTotR54}
\eeq
with $\nu=\pm1$, and hence
\beq
n_\nu=\frac{\rho_\nu}{1-\rho_\nu+\rho_{-\nu}}\,,
\label{n-rho-R54}
\eeq
or, inverting the relation
\beq
\rho_\nu=\frac{1+2 n_{-\nu}}{1+n_\nu+n_{-\nu}}n_\nu\,.
\label{rho-n-R54}
\eeq
The TBA equations for the pseudo-energies read as
\beq
\ep^\nu=\mu^\nu+\ln(1+e^{-\ep^\nu})-\ln(1+e^{-\ep^{-\nu}})=\mu^\nu-\mathtt{T}^\nu_{\,\,\nu'}\ln(1+e^{-\ep^{\nu'}}),
\label{TBAR54}
\eeq
where, and from now on, we omitted the ``$\cdot$'' as we are simply dealing with the summation over two indices. Analogously, we have that the free energy is give by 
\beq
\textgoth{f}=
-\ln(1+e^{-\ep^{+}})-\ln(1+e^{-\ep^{-}})\,.
\eeq
The scattering kernel can also be inferred from~\eqref{rhoTotR54} by noting that
$\rho_{\text{t},\nu}=1_{\text{dr},\nu}$ (or $\rho_\text{t}^{\nu}=1^{\text{dr},\nu}$ if $\rho_\text{t}^{\nu}$ is regarded as a row vector). One can also explicitly work out the $\mathtt{R}$ matrices and their inverses: in terms of the filling functions, they read as
\beq
\begin{split}
\mathtt{R}=&
\begin{pmatrix}
1+n_+ & -n_+\\
-n_-& 1+n_-\\
\end{pmatrix} \quad\mathtt{R}^{-1}=\frac{1}{1+n_++n_-}
\begin{pmatrix}
1+n_- & n_+\\
n_-& 1+n_+\\
\end{pmatrix}, \\
\mathtt{R}^T=&
\begin{pmatrix}
1+n_+ & -n_-\\
-n_+& 1+n_-\\
\end{pmatrix}
\quad\mathtt{R}^{-T}=\frac{1}{1+n_++n_-}
\begin{pmatrix}
1+n_- & n_-\\
n_+& 1+n_+\\
\end{pmatrix}.\\
\end{split}
\label{RMatricesR54}
\eeq
For this model the effective velocities in Eq.~\eqref{FluxJacobian} are found to be~\cite{buca2021rule54}
\beq
v^\text{eff}_\nu=\nu \frac{2}{1+2n_{-\nu}}\,,
\label{vEffR54}
\eeq
and one can easily construct the flux Jacobian using~\eqref{n-rho-R54} and~\eqref{FluxJacobian}, yielding
\beq
\mathtt{A}_\nu^{\,\,\nu'}=\frac{\partial (v^{\text{eff}}_\nu \rho_\nu)}{\partial \rho_{\nu'}}=
\frac{2}{(1+2n_+)(1+2n_-)}
\left(
\begin{array}{cc}
1+2n_+(1+n_{-}) & -2n_+(1+n_{-}) \\
2n_{-}(1+n_+) & -\bigl(1+2n_{-}(1+n_+)\bigr) \\
\end{array}
\right)\,.
\label{FluxJacobianR54}
\eeq
One can easily check that this expression fulfils~\eqref{RARInv} with the definition of the $\mathtt{R}$ matrices given above.

\subsection{Integrable quenches}
\label{IntegrableQuench}

A quantum quench consists in preparing the system in an initial state $|\Omega\rangle$ and letting it to evolve unitarily under the system's Hamiltonian. Even when the latter is integrable, the overlap pattern between the eigenstates of the Hamiltonian and a generic initial state is unstructured and closed-form analytical results are, as a rule, out of reach.
A crucial simplification arises for the class of \emph{integrable} initial states, introduced in Ref.~\cite{ghoshal1994boundary} for integrable quantum field theory and in Ref.~\cite{piroli2017integrable} for integrable lattice models. These are defined as the states that are annihilated by all local conserved charges of the Hamiltonian that are odd under spatial reflection
\beq
Q_{2k+1}\,|\Omega\rangle = 0\,,\qquad k\geq 1\,.
\label{IntCondition}
\eeq
Condition~\eqref{IntCondition} has an immediate and important consequence for the Bethe-ansatz structure of the initial state. Because the eigenvalues $q_{2k+1}(\theta)$ of odd charges are odd functions of the rapidity $\theta$, the constraint~\eqref{IntCondition} forces the overlaps $\langle\{\theta_j\}|\Omega\rangle$ to be non-vanishing only for configurations in which the rapidities come in pairs of opposite sign, $\{\theta_j\}=\{\theta_1,-\theta_1,\theta_2,-\theta_2,\ldots\}$~\cite{piroli2017integrable,piroli2019integrable,pozsgay2014correlations}, where in the multi-index notation $\q=(n,\vartheta)$, $-\theta_i$ means $(n_i,-\vartheta_i)$\footnote{Note, that in models with quasi-particle multiplets, the particle index $n$ must not be identical for a pair.}.
This \emph{pair structure} constitutes the working definition of integrable quenches adopted in this paper:
the initial state is built exclusively from pairs of quasiparticles carrying opposite rapidities (or, equivalently, opposite bare momenta).

\subsubsection{Free Fermions}

For free-fermionic systems the pair structure takes a particularly transparent form.
The initial state can be written as a squeezed-coherent state
\beq
|\Omega\rangle=\mathcal{N}\exp\!\left(\frac{1}{2}\int\frac{\dd\theta}{2\pi}\,
K(\theta)\,a_\theta^\dagger\,a_{-\theta}^\dagger\right)|0\rangle\,,
\label{FFQuenchState}
\eeq
where $a_\theta^\dagger$ creates a fermion with rapidity $\theta$ and the \emph{$K$-function} $K(\theta)$ encodes the quench protocol.
The pair constraint $\rho_\theta=\rho_{-\theta}$ is manifest in the structure of~\eqref{FFQuenchState}, and the logarithmic overlap is simply $\mu(\theta)=-\ln|K(\theta)|^2$.

\subsubsection{Rule 54}

The integrable states for Rule 54 take the form 
\be
|\Omega\rangle= \left(\sqrt{1-n}\ket{00}+ \sqrt{n} \ket{01}\right)^{\otimes L/2}, 
\label{eq:R54QuenchState}
\ee 
where $\{\ket{0},\ket{1}\}$ is the computational basis for one qubit and the parameter $n\in[0,1]$ fixes the quasiparticle occupations. Specifically, recalling the quasiparticles in Rule 54 are specified by the discrete rapidity $\nu=\pm$, after a
quench from a solvable state we have $n_+=n_-=n$~\cite{klobas2021exact2}.

It is worth emphasising that the notion of integrable quench is intrinsically tied to models admitting a quasiparticle description. For certain integrable models---most notably the XYZ spin chain at generic anisotropies---the spectrum is not naturally organised in terms of stable quasiparticle excitations, and the TBA framework recalled in Sec.~\ref{TBA} does not directly apply. The formalism developed in this work is therefore restricted to models with a well-defined quasiparticle content, where the pair-structure condition can be meaningfully imposed.

\section{BMFT for Integrable Quenches}
\label{BMFTIntro}

We want to make statements about the system's dynamics at large, ballistic scales, and at these scales we label fluid cells by the macroscopic coordinates $(x,t)$ which are linked to the microscopic ones by
\beq
\ell x= x_\text{micro}\,,\qquad\qquad \ell t=  t_\text{micro}\,,
\eeq
where $\ell$ is a large scaling parameter. The \emph{Euler scaling limit} is obtained by taking $\ell\to\infty$ at fixed $(x,t)$, while scaling the characteristic wavelength of any inhomogeneity in the state in the same way. Thus, microscopic distances and times grow proportionally, with the ballistic rays $x/t=x_{\rm micro}/t_{\rm micro}$ kept fixed. Observables are understood as averages over mesoscopic fluid cells of size
\beq
\xi_{\rm micro}\ll \ell_{\rm cell}\ll\ell\,
\eeq
so that each cell is locally homogeneous, while thermodynamic parameters vary smoothly on the scale $\ell$. At this scale only the leading, first-gradient conservation laws are retained, whereas diffusive and higher-gradient corrections are subleading.

Before specialising to the quench setting, let us briefly recall the structure of ballistic macroscopic fluctuation theory (BMFT) as formulated in Refs.~\cite{doyon2023bmft,doyon2023longrange}. BMFT is a universal framework for describing fluctuations and correlations at the Euler scale in many-body systems supporting ballistic transport.

It rests on two independent pillars:
\begin{enumerate}
\item \emph{The generalised Boltzmann--Gibbs principle.}
At the Euler scale the only relevant degrees of freedom are the conserved-density fields, and their probability is entirely governed by the thermodynamic entropy.
This is not a statement restricted to the initial time: it holds throughout the evolution, reflecting the fact that at every macroscopic space-time point the system is locally described by a (possibly inhomogeneous) generalised Gibbs ensemble (GGE)~\cite{doyon2023bmft}.

```latex
\item \emph{Noiseless, deterministic Euler evolution.}
At the Euler scale, each realization of the fluctuating hydrodynamic fields is transported ballistically according to the GHD continuity equation
\beq
\partial_t \rho_\q(x,t)
+\partial_x\!\left(v^{\mathrm{eff}}_\q(x,t)\rho_\q(x,t)\right)=0\,.
\label{BMFTContinuity}
\eeq
No stochastic forcing appears at this scale: once the initial profile is fixed, its subsequent evolution is completely determined by the nonlinear Euler equations, including the state dependence of the effective velocity \(v^{\mathrm{eff}}_\q\).
\end{enumerate}
Thus, all stochasticity in BMFT enters through the initial condition, and the full statistical content of the theory is encoded in the probability measure over the initial conserved-density profiles. This noiseless description is supported by a broad range of results on diffusive corrections and hydrodynamic fluctuations in integrable systems
\cite{doyon2026noise,Hubner_2025,medenjak2020diffusion,
denardis2019diffusion,denardis2018hydrodynamic,
ilievski2018superdiffusion,gopalakrishnan2018operator,
medenjak2017lower}.
The initial measure is therefore a central ingredient of the construction. For an initial state described by a (possibly inhomogeneous) GGE with chemical potentials $\mu^i(x)$, the measure is inferred from the thermodynamic properties of the initial state via standard large-deviation arguments (and can be verified to reproduce all Euler-scale connected correlation functions of conserved densities).
It takes the form
\beq
\dd \mathbb{P}[\{\mathsf{q}_{i,t=0}\}]=\frac{1}{Z}\,\dd \mathcal{M}[\{\mathsf{q}_{i,t=0}\}]\,
e^{-\ell\,\mathcal{F}[\{\mathsf{q}_{i,t=0}\}]}\,,
\label{BMFTInitMeasure}
\eeq
where the rate functional is given by 
\beq
\mathcal{F}[\{\mathsf{q}_{i,t=0}\}]=\int \dd x\left[\mu^i(x)\,\mathsf{q}_i(x,0)-s[\{\mathsf{q}_i(x,0)\}]-\textgoth{f}(\{\mu^i(x)\})\right]\,,
\label{BMFTRateFunctional}
\eeq
with $s[\{\mathsf{q}_i\}]$ the thermodynamic entropy density evaluated at the density profile $\{\mathsf{q}_i\}$ and $\textgoth{f}(\{\mu^i\})$ the equilibrium free-energy density at chemical potentials $\{\mu^i\}$. In the previous expressions and the formula above, the curly brackets stress the fact that the entropy density is a function of all local densities, and similarly, the free energy-density is a function of all chemical potential indexed by $i$.
The functional $\mathcal{F}$ penalises deviations of $\mathsf{q}_{i}(x,0)$ from the local-equilibrium profile exponentially in $\ell$.
The time-dependent measure is obtained by restricting~\eqref{BMFTInitMeasure} to trajectories satisfying~\eqref{BMFTContinuity}, which is enforced through some Lagrange-multiplier fields that we denote by $H(x,t)$.

In practice, all Euler-scale fluctuations and correlations---including the full counting statistics (FCS) of transported charges, dynamical two-point functions, and the emergence of long-range spatial correlations from inhomogeneous initial conditions~\cite{doyon2023longrange}---are thereby reduced to a deterministic saddle-point problem in the $\ell\to\infty$ limit.
The approach was originally applied to inhomogeneous GGE initial states, both in the general hydrodynamic setting~\cite{doyon2023bmft} and within GHD for interacting integrable models~\cite{doyon2020fluctuations,perfetto2021euler}.
More recently it has been used to compute anomalous current fluctuations in cellular automata~\cite{yoshimura2025anomalous}, and its foundations have been further elucidated through a mapping of interacting integrable models to ensembles of free point particles~\cite{kethepalli2025bmft}.

In this setting we now propose a path-integral treatment, completely analogous to BMFT, to describe leading-order evolution after integrable quenches.
The key new ingredient is the pair structure of the initial state (cf.\ Sec.~\ref{IntegrableQuench}), which modifies the initial measure while the evolution prescription~\eqref{BMFTContinuity} remains unchanged.

\subsection{The initial measure}
\label{sec:initial_measure}

BMFT is conventionally formulated in terms of the fluctuating conserved-charge densities $\mathsf{q}_i(x,0)$. In an integrable system, however, a local thermodynamic state can equivalently be parametrised by its root densities $\rho_\q(x,0)$; for a complete set of charges, this amounts to a change of coordinates on the space of local thermodynamic states. This parametrisation is particularly natural for our purposes, since root densities are also the variables underlying the Quench Action approach. We therefore adopt them as dynamical variables from the outset.

We recall~\eqref{RapidityDepMu} and the relation between the root and charge densities,
\beq
\mu^\q=\mu^i \cdot q_i^\q\,,\quad \text{and}\quad \mathsf{q}_i=q_i^\q \cdot \rho_\q\,,
\label{RapidityChargeDensityDepMu}
\eeq
which imply
\beq
\mu^i \, \mathsf{q}_i=\mu^\q \cdot \rho_\q\,.
\eeq
Thus, the thermodynamic pairing entering the BMFT rate functional is unchanged by this reparametrisation. Having expressed the BMFT measure directly in terms of root-density profiles, its initial-state weight can now be inferred from the Quench Action approach~\cite{caux2013quenchaction,caux2016quenchaction,essler2016quench}.

The starting point is the diagonal ensemble, which governs the long-time behaviour of local observables after a quantum quench from a homogeneous initial state $|\Omega\rangle$.
In the thermodynamic limit the trace over energy eigenstates can be reorganised as a functional integral over root-density profiles $\rho_\q$: distinct microstates sharing the same macroscopic root density are grouped together, contributing the Yang--Yang entropy $e^{L\,s[\rho]}$ as a density of states, while their overlap with the initial state contributes the weight $|\langle \rho|\Omega\rangle|^2$~\cite{caux2016quenchaction,pozsgay2011meanvalues}.
For an integrable quench (cf.\ Sec.~\ref{IntegrableQuench}), the pair structure ensures that the extensive part of the logarithmic overlap takes the form~\cite{piroli2017integrable,caux2016quenchaction}
\beq
-\ln|\langle\{\q_j,-\q_j\}|\Omega\rangle|^2\;\sim\; L\,\mu^\q\cdot\rho_\q\,,
\label{LogOverlap}
\eeq
where $\mu^\q=\mu^{-\q}$ is the \emph{overlap function} encoding the quench protocol, and it is understood as the rapidity dependent chemical potentials \eqref{RapidityChargeDensityDepMu}.
The integration runs over all rapidities; however, because each pair $(\q,-\q)$ is counted once, the natural weight in the diagonal ensemble is
\beq
\exp\!\Big[-L\,\mathcal{F}_\text{QQ}[\rho]\Big]\,,\qquad
\mathcal{F}_\text{QQ}[\rho]=\frac{1}{2}\left\{\frac{\mu^\q\cdot\rho_\q}{2\pi}\,\;-\;s[\rho]\right\}\,,
\label{QuenchActionFunctional}
\eeq
where the factor $\tfrac{1}{2}$ reflects the pair structure.
Extremising $\mathcal{F}_\text{QQ}$ yields a set of generalised TBA equations whose solution is the saddle-point root density $\bar\rho_\q$; this coincides with the GGE describing the post-quench steady state~\cite{caux2013quenchaction,caux2016quenchaction}.

To pass from the diagonal ensemble to a probability measure suitable for BMFT we promote $\rho_\q$ to a spatially dependent field $\rho_\q(x,0)$, following the same logic as in the standard BMFT construction (cf.\ the beginning of Sec.~\ref{BMFTIntro}).
The quench action functional~\eqref{QuenchActionFunctional} then provides the large-deviation rate function for the initial root-density fluctuations.
Putting everything together, we obtain the probability measure
\beq
\dd \mathbb{P}_\text{QQ}[\rho_{t=0}]=\frac{1}{Z}\dd \mathcal{M}^*[\rho_{t=0}]\,e^{-\ell\,\mathcal{F}_\text{QQ}[\rho_{t=0}]}\,,
\label{ConstrainedMeasure}
\eeq
where the the superscript $\ast$ indicates that fluctuating fields are constrained by the pair structure
\beq
\rho_{\q}(x,0)=\rho_{-\q}(x,0)\,.
\eeq
For the sake of normalisation and to make $\mathcal{F}_\text{QQ}$ manifestly non-negative, we redefine the rate functional by subtracting its value at the saddle point
\beq
\mathcal{F}_\text{QQ}[\rho_{t=0}]=\frac{1}{2}\int \dd x \left[\mu^\q \cdot \rho_\q(x,0)- s[\rho_{t=0}]-\textgoth{f}_\text{GGE}(\mu)\right]\,,
\eeq
where $\mu^\q$ is the (homogeneous) overlap function and $\textgoth{f}_\text{GGE}(\mu)$ is the free-energy density of the post-quench steady-state GGE. The above quantities generally depend on a (n infnite) set of chemical potentials. To lighten the notations, when the parameteres are rapidity-dependent, we shall refer to a collection of them without using curly brackets or other vecor-notations. That is $\mu$ and $\rho$ stand for $\{\mu^\q\}$ and $\{\rho_\q\}$. For quantities with latin indices, we shall keep the $\{\mu^i\}$ notation. Notably, this function is entirely characterised by the thermodynamic data of the late-time steady state and, in fact, $\mathcal{F}_\text{QQ}=1/2\, \mathcal{F}_\text{GGE}$.

Before proceeding, let us briefly discuss the normalisation factor \(Z\).  By the TBA variational principle we have 
\begin{equation}
  \mu^\q\cdot\rho_\q-s[\rho]
  -\textgoth{f}_{\rm GGE}(\mu)
  \geq0,
  \label{eq:TBA-rate-nonnegative}
\end{equation}
with equality at the homogeneous GGE root density
\(\rho_\q=\bar\rho_\q\). Namely we can write 
\begin{equation}
  \mathcal F_{\rm QQ}[\rho]\geq0,
  \qquad
  \mathcal F_{\rm QQ}[\bar\rho]=0.
  \label{eq:QQ-rate-minimum}
\end{equation}
The homogeneous profile is compatible with the pair constraint and generates
a stationary homogeneous solution of the GHD equation.  It therefore remains
an admissible configuration after both constraints are imposed.  Since
imposing constraints only restricts the variational domain, the constrained
minimum of \(\mathcal F_{\rm QQ}\) is still zero. The Laplace principle then
gives
\begin{equation}
  \lim_{\ell\to\infty}\frac{1}{\ell}\log Z=0.
  \label{eq:normalisation-subextensive}
\end{equation}
Thus the normalisation contributes no leading extensive term.

The pair-constrained measure in Eq.~\eqref{ConstrainedMeasure} can be written in terms of an unconstrained one by introducing an auxiliary field (or Lagrange multiplier) as follows
\beq
\begin{split}
\dd \mathbb{P}_\text{QQ}[\rho_{t=0}]&=\frac{1}{Z}\dd \mathcal{M}[\rho_{t=0}]\,e^{-\ell\,\mathcal{F}_\text{QQ}[\rho_{t=0}]}\,\delta[\rho_\q(x,0)-\rho_{-\q}(x,0)]\\
&=\frac{1}{Z}\dd \mathcal{M}[\rho_{t=0}]\,\dd \mathcal{M}[P]\,\exp\!\left[{-\ell\,\mathcal{F}_\text{QQ}[\rho_{t=0}]-\frac{\ell}{2} \int \dd x \, P^\q(x)\cdot\bigl(\rho_\q(x,0)-\rho_{-\q}(x,0)\bigr)}\right]\,.
\end{split}
\label{QQ_BMFT_Init_Measure}
\eeq
Note the factor in front of the second term of the exponential is arbitrary and we chose $1/2$ for convenience (one can always modify this choice by performing a re-scaling of $P_\theta$). Instead, the factor $(\rho_\q(x,0)-\rho_{-\q}(x,0))$ inside the integral is fixed. For example, it cannot be replaced by $(\rho_\q(x,0)-\rho_{-\q}(x,0))^m$ with $\mathbb N \ni m>1$ as this would not satisfy the Lagrange multiplier theorem. 

It is easy to see that the FCS of conserved charges in the initial state, characterised by the measure above, is
\beq
f_0(\lambda,\mu)= \frac{1}{2}\left(\textgoth{f}_\text{GGE} (\mu)-\textgoth{f}_\text{GGE} (\mu-2\lambda q_i)\right)= \frac{1}{2}\left(\textgoth{f}_\text{GGE} (\{\mu^j\})-\textgoth{f}_\text{GGE} (\{\mu^j-2\lambda\delta^{i,j}\})\right)=\frac{1}{2}f_\text{GGE}(2\lambda,\mu)\,,
\eeq
where the $X$ dependence drops out for inhomogeneous states in the $\ell\rightarrow \infty$ limit. This result, linking the fluctuation in integrable initial states and the corresponding GGEs, is consistent with the findings of Refs.~\cite{horvath2024fcs,bertini2024dynamics}, where the initial-state FCS was computed independently, thus providing a non-trivial check of the proposed measure.

This relation between the initial-state and GGE fluctuations ~\cite{horvath2024fcs} is understood as a functional identity for conserved charges whose initial FCS has a nontrivial extensive limit.  It was derived using Quench Action techniques in Ref.~\cite{horvath2024fcs}, although the thermodynamic steps involved are not fully rigorous and the derivation is not independent of the same Quench Action structure underlying Eq.~\eqref{QQ_BMFT_Init_Measure}.  Importantly, the relation admits more direct and stronger verification in several cases.  For free-fermionic, integrable initial states and for the Bose--Einstein-condensate initial state in the Lieb--Liniger model it follows from explicit microscopic calculations. Moreover, the first few cumulants were also tested numerically in interacting examples in Ref.~\cite{horvath2024fcs}.  %The free-fermionic time-dependent result is furthermore consistent with Ref.~\cite{bertini2024dynamics}.

These results support the pair interpretation of the factors \(2\lambda\) and \(1/2\): initially the two members of a correlated pair contribute coherently to the charge, while paired Bethe configurations carry half of the unconstrained thermodynamic entropy.

Together, these analytical and numerical results provide strong support for the functional relation in fluctuating charge sectors.  Work in progress aims to place its derivation on firmer mathematical ground, in particular through boundary-splitting and extensivity results controlling the subsystem thermodynamic limit for dimer states, with possible extensions to suitable matrix product states.

We also note that the measure in Eq.~\eqref{QQ_BMFT_Init_Measure}, in its present form, presupposes an extensive and non-degenerate large-deviation structure in the charge directions retained as fluctuating hydrodynamic fields.  This need not hold for every conserved charge: N\'eel and dimer states can, for example, be exact eigenstates of magnetisation or particle number, even though their energy and generic higher parity-even charges fluctuate extensively.

An exactly fixed charge does not necessarily destroy the large-deviation principle, but makes the joint fluctuation measure singular and restricts it to a lower-dimensional manifold of root-density profiles.  In a multi-species Bethe-ansatz system this restriction can couple different string densities and may modify the fluctuations of the remaining charges. We therefore do not assume that every fixed direction can be incorporated by appending an independent delta-functional constraint to an otherwise unchanged measure. As argued in Appendix~\ref{app:overlap}, such non-extensive directions are nevertheless expected to be exceptional for finite-range parity-even charges in dimer-type integrable initial states.  We apply \eqref{QQ_BMFT_Init_Measure} only in directions with a nontrivial extensive initial FCS; symmetry-fixed sectors and their coupling to other Bethe-root fluctuations are left for future work.

\subsection{The BMFT-QQ functional for arbitrary times}
\label{sec:BMFT_QQ_functional}

Proceeding along the lines of the general BMFT construction, to compute quantities at finite time we impose our dynamical variables $\rho_{\q}$ to fulfil the GHD continuity equation~\eqref{BMFTContinuity}. Following Refs.~\cite{doyon2023bmft,doyon2023longrange} the constraint is implemented by introducing a Lagrange-multiplier field $H^\q(x,t)$. Namely, combined with the initial measure~\eqref{QQ_BMFT_Init_Measure}, the full time-dependent measure reads
\beq
\begin{split}
\dd \mathbb{P}_\text{QQ}[\rho]&=\dd \mathbb{P}_\text{QQ}[\rho_{t=0}]\,\delta[\partial_t\rho_\q(x,t)+\partial_x(v^\text{eff}_\q(x,t)\,\rho_\q(x,t))]\,\delta[\rho_\q(x,0)-\rho_{-\q}(x,0)]\\
&=\frac{1}{Z}\dd \mathcal{M}[\rho]\,\dd \mathcal{M}[H]\,\dd \mathcal{M}[P]\,\exp\!\left[{-\ell\,\mathcal{F}_\text{QQ}[\rho_{t=0}]-\frac{\ell}{2} \int \dd x \, P^\q(x)\cdot\bigl(\rho_\q(x,0)-\rho_{-\q}(x,0)\bigr)}\right]\\
&\qquad\qquad\qquad\qquad\qquad\times\exp\!\left[{-\frac{\ell}{2}\int \dd x \int_0^{T+0^+} \hspace{-0.5cm}\dd t \,H^\q \cdot \bigl(\partial_t \rho_\q+\partial_x (v^\text{eff}_\q\,\rho_\q)\bigr)}\right]\,.
\end{split}
\label{QQ_BMFT_TimeDep_Measure}
\eeq
Note that we again chose the arbitrary factor in front of the third exponent to be $1/2$. We emphasise that, when seeking for saddle-point solutions of the action defined by~\eqref{QQ_BMFT_TimeDep_Measure}, one must perform the variation also with respect to $H^\q$ and $P^\q$.

%Before proceeding, let us briefly address the normalisation factor $Z$. For $\ell\to\infty$, the path integral is dominated by the saddle point. For the \emph{unbiased} functional (i.e.\ without the insertion of any observable), the saddle-point configuration is the trivial one: $H^\q(x,t)=0$, $P^\q(x)=0$, and the root density takes its homogeneous GGE value $\rho_\q(x,0)=\bar\rho_\q$ everywhere. \bruno{In principle you could have other solutions, no? You need to argue that this is the leading one. This would be a lot easier if the terms in the exponent multiplying the auxiliary fields were positive.}{\color{blue}D: This solution is unique, at least on an appropriate space of functions. Cf.Eq. 57 with no inhomogeneity for H, but (H(T)==0).} With this configuration, every term in the exponent of~\eqref{QQ_BMFT_TimeDep_Measure} vanishes identically---the rate functional $\mathcal{F}_\text{QQ}$ vanishes at its saddle point by construction, the continuity equation is trivially satisfied, and the pair constraint holds.
%Therefore $Z=1$ at leading order in $\ell$, and the normalisation factor can be set to unity throughout the subsequent analysis.

With the time-dependent measure, we can evaluate the expectation value of any ``mesoscopic'' observable by inserting the appropriate functional into the path integral~\eqref{QQ_BMFT_TimeDep_Measure}.
A central consequence of the generalised Boltzmann--Gibbs principle (cf.\ the beginning of Sec.~\ref{BMFTIntro}) is that, at the Euler scale, mesoscopic (fluid-cell averaged) observables do not fluctuate independently from the conserved densities: they are instead determined, in general nonlinearly, by the local values of the fluctuating root densities~\cite{doyon2023bmft}.
Concretely, the fluid-cell mean $\bar{o}(\ell x,\ell t)$ of a local observable $\hat{o}$ is identified with its GGE average evaluated in the local state specified by the fluctuating root densities,
\beq
\bar{o}(\ell x,\ell t)\;\equiv\;\mathtt{o}[\rho(x,t)]\qquad\text{(at Euler scale)}\,,
\label{BGPrinciple}
\eeq
where $\mathtt{o}[\rho]=\langle \hat{o}\rangle_\text{GGE}|_{\rho}$.
This identification extends to multipoint correlation functions: the Euler-scale connected $n$-point function of observables $\hat{o}_1,\ldots,\hat{o}_n$ at macroscopically separated space-time points is obtained as~\cite{doyon2023bmft}
\beq
S_{\hat{o}_1,\ldots,\hat{o}_n}(x_1,t_1;\ldots;x_n,t_n)
=\lim_{\ell\to\infty}\ell^{n-1}\,
\bigl\langle \mathtt{o}_1[\rho(x_1,t_1)]\cdots \mathtt{o}_n[\rho(x_n,t_n)]\bigr\rangle^{\rm c}_\text{QQ}\,,
\label{EulerCorrelators}
\eeq
where $\langle\cdots\rangle_\text{QQ}$ denotes the average with respect to the measure~\eqref{QQ_BMFT_TimeDep_Measure} with $T \ge \text{max}\, t_i$, and the superscript ``c'' indicates the connected part.
The prefactor $\ell^{n-1}$ compensates the $\ell^{-(n-1)}$ scaling of the connected $n$-point function of macroscopic variables, ensuring a finite limit.
In the $\ell\to\infty$ limit, the path integral is dominated by its saddle point, and the computation of~\eqref{EulerCorrelators} reduces to solving the associated saddle-point equations, possibly with source terms representing the observable insertions.

Turning to the FCS, we note that its Euler scaling is simpler than that of pointwise correlation functions. Let
\beq
\widehat{Q}_{i,X}^{(\ell)}(T)
=\int_{-\ell X/2}^{\ell X/2}\dd x_{\rm micro}\,
\hat q_i(x_{\rm micro},\ell T)
\label{EulerCharge}
\eeq
be the charge contained in a macroscopic interval of length $\ell X$. At the Euler scale, this observable is represented in BMFT by the functional of the fluctuating hydrodynamic density
\beq
Q_{i,X}^{(\ell)}[\mathsf{q}](T)
:=
\ell\int_{-X/2}^{X/2}\dd x\,\mathsf{q}_i(x,T)\,.
\label{EulerChargeBMFT}
\eeq
Accordingly,
\beq
\left\langle e^{\lambda\widehat{Q}_{i,X}^{(\ell)}(T)}\right\rangle_\ell
\underset{\rm Euler}{\asymp}
\left\langle
\exp\,\left[
\lambda\ell\int_{-X/2}^{X/2}\dd x\,\mathsf{q}_i(x,T)
\right]
\right\rangle_{\rm QQ}\,,
\label{EulerChargeMGF}
\eeq
and the Euler-scale FCS, or scaled cumulant generating function, is defined by
\beq
f(\lambda,\mu,X,T)=\lim_{\ell\to\infty}\frac{1}{\ell X}
\ln\left\langle
e^{\lambda\widehat{Q}_{i,X}^{(\ell)}(T)}
\right\rangle_\ell=\lim_{\ell\to\infty}\frac{1}{\ell X}
\ln\left\langle
\exp\,\left[
\lambda\ell\int_{-X/2}^{X/2}\dd x\,\mathsf{q}_i(x,T)
\right]
\right\rangle_{\rm QQ}\,,
\label{EulerFCSDefinition}
\eeq
where the reference to the physical charge $i$ is implicit. No additional fluid-cell averaging is required here: the integral over a macroscopic region already coarse-grains the local density, and replacing it by its fluid-cell mean does not affect the leading Euler-scale result~\cite{doyon2023bmft}. In contrast to connected $n$-point functions at macroscopically separated points, which scale as $\ell^{1-n}$ and require the compensating prefactor in~\eqref{EulerCorrelators}, the logarithm of the moment generating function is extensive, namely of order $\ell$. Consequently, after the sole extensive normalisation $1/(\ell X)$, the FCS in Eq.~\eqref{EulerFCSDefinition} has a finite and generically non-vanishing limit, without the additional compensating factor required for pointwise connected correlation functions.

\section{The time-dependent FCS after integrable quenches}
\label{sec:time_dep_FCS}

We now compute the FCS of a conserved quantity after an integrable quench using the path integral defined above. The scaled cumulant generating function is
\beq
\begin{split}
f(\lambda,\mu, X,T)&=\lim_{\ell\rightarrow \infty} (X\ell)^{-1}\ln \left[\langle e^{\ell \lambda\int_{-X/2}^{X/2} \dd x\, q_i^\q \cdot\rho_\q (x,T)}\rangle_\text{QQ}\right]\\
&=\lim_{\ell\rightarrow \infty} (X\ell)^{-1}\ln \left[\int  \dd \mathcal{M}[\rho,H,P]\,e^{-\ell\mathcal{F}[\rho_{t=0}]-\ell\int \dd x \int_0^T \dd t\, H^\q \cdot (\partial_t \rho_\q+\partial_x (v^\text{eff}_\q\rho_\q))}  \right.\\
&\qquad \qquad  \qquad \qquad \quad \left. \times\, e^{-\frac{\ell}{2}  \int \dd x \, P^\q(x)\cdot(\rho_\q(x,0)-\rho_{-\q}(x,0))}\,e^{\frac{\ell}{2} \cdot 2\lambda\int_{-X/2}^{X/2} \dd x\, q_i^\q \cdot\rho_\q (x,T) }\right]\\
&\equiv \lim_{\ell\rightarrow \infty} (X\ell)^{-1}\ln \left[\int  \dd \mathcal{M}[\rho,H,P]\,e^{-\ell\,\mathcal{F}_\text{QQ-FCS}[\rho,H,P]}\right]\,,
\end{split}
\eeq
where we used that $\lim_{\ell \rightarrow \infty} Z=1$,  we set $\dd \mathcal{M}[\rho,H,P]\equiv\dd \mathcal{M}[\rho]\,\dd \mathcal{M}[H]\,\dd \mathcal{M}[P]$ for brevity, and we defined
\beq
\begin{split}
\mathcal{F}_\text{QQ-FCS}[\rho,H,P]&\equiv \frac{1}{2}\left[\int \dd x \,\mu^\q \cdot \rho_\q(x,0)- s[\rho(x,0)]-\textgoth{f}_\text{GGE}(\mu)\right]+\frac{1}{2}\int \dd x \int_0^{T+0^+} \dd t\, H^\q \cdot (\partial_t \rho_\q+\partial_x (v^\text{eff}_\q \rho_\q))\\
&\qquad+\frac{1}{2}  \int \dd x \, P^\q(x) \cdot(\rho_\q(x,0)-\rho_{-\q}(x,0))-  \frac{1}{2}\int_{-X/2}^{X/2} \dd x\, 2\lambda\, q_i^\q \cdot\rho_\q (x,T)\,.
\end{split}
\label{FCS_QQ_Functional}
\eeq
For $\ell\rightarrow \infty$, the path integral is dominated by its saddle point, which is described by the following equations 
\beq
\begin{split}
&0=\frac{\delta \mathcal{F}_\text{QQ-FCS}}{\delta \rho_\q(x,t)}=-\partial_t H^\q(x,t)-\mathtt{A}_\phi^{\,\,\,\q} \partial_x H^\phi(x,t)-2\lambda \delta(t-T)\,\text{B}_{X/2}(x)q_i^\q\\
&\qquad\qquad\qquad\quad+\delta(t)\left(\mu^\q-\beta^\q[\rho](x,0)-H^\q(x,0)+P^\q(x)-P^{-\q}(x)\right)\\
&\qquad\qquad\qquad\quad-\delta(t-T-0^+)H^\q(x,T+0^+),\\
&0=\frac{\delta \mathcal{F}_\text{QQ-FCS}}{\delta H^\q(x,t)}=\partial_t\rho_\q(x,t)+\partial_x(v^\text{eff}_\q \rho_\q(x,t)),\\
&0=\frac{\delta \mathcal{F}_\text{QQ-FCS}}{\delta P^\q(x,0)}=\rho_\q(x,0)-\rho_{-\q}(x,0),
\label{SP1_QQ-FCS_B}
\end{split}
\eeq
for any $x\in \R$ and $t\in [0,T+0^+]$. Here we introduced the box function $\text{B}_{X/2}(x)$, equal to~$1$ inside the interval $[-X/2,X/2]$ and zero elsewhere, and we encountered the chemical potentials $\beta^\q$ exploiting that in equilibrium, $\rho$ and $\beta$ are linked by
\beq 
\beta^\q=\frac{\dd s[\rho]}{\dd \rho_\q}%\right|_{\rho_\q=\mathtt{\rho}_\q(\beta)}
=\ln\left(\frac{1-n^\q}{n^\q}\right)-\ln\left(1-n^\phi\right)\cdot\mathtt{T}_\phi^{\,\,\,\q}\,,\quad\text{and}\quad
\frac{\dd \,\textgoth{f}(\beta)}{\dd \beta^\q}=-\mathtt{\rho}_\q(\beta)\,,
\label{eq:betavsn}
\eeq
for a given free energy and entropy density $\textgoth{f}$ and $s$. 
In particular, $\beta^\q[\rho]$ means the set of chemical potentials that describe the same macro-state as the given root density $\rho$ and they are deliberately denoted by $\beta^\q$ to distinguish them from the chemical potentials $\mu^\q$ that are associated with the steady-state GGE. The $\beta$ and $\rho$ quantities are related by dressing and TBA equations Eqs. \eqref{rho-rhoTot},\eqref{TBAEq} and \eqref{PseudoE},\eqref{FillingFunction}; and the space-time dependence is naturally attributed to these objects in the fluid-cell picture. For simplicity, we denote $\beta^\q[\rho(x,t)]$ by $\beta^\q[\rho](x,t)$.
Note that the boundary terms in the second line of the first equation~\eqref{SP1_QQ-FCS_B} arise from the initial constraint, i.e.\ from $\mathcal{F}_\text{QQ}$, and from integration by parts transferring the time derivative from $\rho$ to $H$. These boundary terms can be treated separately by integrating in time respectively around $t=0$ and $t=T$. This gives  
\beq
\begin{split}
&\mu^\q-\beta^\q[\rho](x,0)-H^\q(x,0)+P^\q(x)-P^{-\q}(x)=0,\\
&H^\q(x,T+0^+)=0\,.
\label{SP2_QQ-FCS_B}
\end{split}
\eeq
We now combine the last equation of~\eqref{SP1_QQ-FCS_B} and the first one of~\eqref{SP2_QQ-FCS_B} using the following argument. If $\rho_\q (x,0)=\rho_{-\q}(x,0)$, i.e.\ the root density is symmetric under rapidity inversion at fixed~$x$, then the TBA equations relating $\rho_\q$ and $\beta^\q$ guarantee that $\beta^\q(x,0)=\beta^{-\q}(x,0)$ as well.
This means that we can write
\beq
\begin{split}
\beta^{\q}(x,0)&=\mu^\q-H^\q(x,0)+P^\q(x)-P^{-\q}(x)\\
&=\mu^\q-H^{-\q}(x,0)+P^{-\q}(x)-P^{\q}(x)\,,\\
\end{split}
\eeq
taking the difference we find 
\beq
P^\q(x)-P^{-\q}(x)=\frac{1}{2}\left(H^\q(x,0)-H^{-\q}(x,0)\right)\,,
\eeq
and hence~\footnote{Note that this equation implies that the symmetric part of $P^\q$ is undetermined, however, it does not affect the action at the saddle point.}
\beq
\beta^{\q}(x,0)=\beta^{-\q}(x,0)=\mu^\q-\frac{1}{2}\left(H^\q(x,0)+H^{-\q}(x,0)\right)\,.
\eeq
To summarise, we find that the saddle-point equations can be reduced to 
\beq
\begin{split}
&\partial_t H^\q(x,t)+\mathtt{A}_\phi^{\,\,\,\q} \partial_x H^\phi(x,t)=-2\lambda \delta(t-T)\,\text{B}_{X/2}(x)q_i^\q,\\
&\partial_t\rho_\q(x,t)+\partial_x(v^\text{eff}_\q \rho_\q(x,t))=0,\\
&\beta^{\q}(x,0)=\beta^{-\q}(x,0)=\mu^\q-\frac{1}{2}\left(H^\q(x,0)+H^{-\q}(x,0)\right),\\
&H^\q(x,T+0^+)=0\,,
\label{SP_QQ-FCS_WithH_B}
\end{split}
\eeq
where we used that $\mu^\q=\mu^{-\q}$ and $q_i^\q=q_i^{-\q}$.
For $t \in [0,T]$, these equations can be rewritten by absorbing the delta-function inhomogeneity into an explicit terminal condition for the $H$ field
\beq
\begin{split}
&\partial_t H^\q(x,t)+\mathtt{A}_\phi^{\,\,\,\q} \partial_x H^\phi(x,t)=0,\\
&\partial_t\rho_\q(x,t)+\partial_x(v^\text{eff}_\q \rho_\q(x,t))=0,\\
&\beta^{\q}(x,0)=\beta^{-\q}(x,0)=\mu^\q-\frac{1}{2}\left(H^\q(x,0)+H^{-\q}(x,0)\right),\\
&H^\q(x,T)=2\lambda \,\text{B}_{X/2}(x)q_i^\q\,.
\label{SP_QQ-FCS_WithH_C}
\end{split}
\eeq
Once the saddle-point solution for $\rho_\q$ and $H^\q$ are found from~\eqref{SP_QQ-FCS_WithH_C}, the FCS can be expressed as
\beq
\begin{split}
f(\lambda,\mu, X,T)&=-X^{-1} \mathcal{F}_\text{QQ-FCS}[\rho_\q,H^\q]\\
&=-\frac{1}{2X}\left\{\left[\int \dd x \,\mu^\q \cdot \rho_\q(x,0)- s[\rho(x,0)]-\textgoth{f}_\text{QQ}(\mu)\right]- \int \dd x\, 2\lambda\, q_i^\q \cdot\rho_\q (x,T) \,\text{B}_{X/2}(x)\right\}\\
&=-\frac{1}{2X} \left\{\left[\int \dd x \,\textgoth{f}_\text{QQ}({\beta}(x,0))-\textgoth{f}_\text{QQ}(\mu)\right]-\int \dd x\, (\beta^\q(x,0)-\mu^\q)\cdot\rho_\q (x,0)\right.\\
&\qquad\qquad\quad\left.- \int \dd x\, 2\lambda\, q_i^\q \cdot\rho_\q (x,T)\,\text{B}_{X/2}(x)\right\}\,,
\end{split}
\label{NonEqFCSWithSpSol}
\eeq
where in the last step we used that $\beta^\q(x,0)$ and $\rho_\q(x,0)$ are linked by the TBA equations~\eqref{TBAEq} (together with~\eqref{rho-rhoTot} and~\eqref{FillingFunction}).

Importantly, by rescaling the coordinates as $x\to x/X$ and $t\to t/X$, one can easily show that the FCS in Eq.~\eqref{NonEqFCSWithSpSol} only depends on $T$ and $X$ only through the ratio
\beq
\zeta=\frac{T}{X}\,,
\eeq
as expected in the Euler scaling limit. More explicitly, introducing the rescaled functions 
\be
\label{eq:rescaledfuncts}
\begin{aligned}
\tilde H^\q(x,t) & = H^\q(x X,t X), & \tilde \beta^\q(x,t) & =\beta^\q(x X,t X), &  \tilde\rho_\q(x,t) &=\rho_\q(x X,t X), \\
\tilde v^\text{eff}_\q(x,t) & =v^\text{eff}_\q(x X,t X), &  {\mathtt{A}}_\phi^{\,\,\,\q} (x,t) & = {\mathtt{\tilde A}}_\phi^{\,\,\,\q} (x X,t X), 
\end{aligned}
\ee
we have that they fulfil 
\beq
\begin{split}
&\partial_t \tilde H^\q(x,t)+ {\mathtt{\tilde A}}_\phi^{\,\,\,\q} \partial_x \tilde H^\phi(x,t)=0,\\
&\partial_t \tilde \rho_\q(x,t)+\partial_x(\tilde v^\text{eff}_\q \tilde \rho_\q(x,t))=0,\\
&\tilde \beta^{\q}(x,0)= \tilde \beta^{-\q}(x,0)=\mu^\q-\frac{1}{2}\left(\tilde H^\q(x,0)+\tilde H^{-\q}(x,0)\right),\\
&\tilde H^\q(x,\zeta)=2\lambda \,\text{B}_{1/2}(x){q}_i^\q\,.
\label{eq:SPrescaled}
\end{split}
\eeq
Therefore they do indeed only depend on $T$ and $X$ through their ratio. Changing now variables as $x\to x'=x/X$ and $t\to t'=t/X$ in the integrals in Eq.~\eqref{NonEqFCSWithSpSol} we have 
\be
\begin{aligned}
f(\lambda,\mu, X,T) &=-\frac{1}{2} \left\{\left[\int \dd x' \,\textgoth{f}_\text{QQ}({\tilde \beta}(x',0))-\textgoth{f}_\text{QQ}(\mu)\right]-\int \dd x'\, (\tilde \beta^\q(x',0)-\mu^\q)\cdot\tilde\rho_\q (x',0)\right.\\
&\qquad\qquad\quad\left.- \int \dd x'\, 2\lambda\, q_i^\q \cdot\tilde\rho_\q (x',\zeta)\,\text{B}_{1/2}(x')\right\} \equiv f(\lambda,\mu, \zeta). 
\end{aligned}
\label{eq:Frescaled}
\ee
Although they represent an immense simplification compared to solving the full many-body dynamics, Eqs.~\eqref{eq:SPrescaled} are still very difficult to solve, both analytically and numerically. Here we will present explicit solutions in two important cases, which we sort in order of increasing difficulty: (i) free fermionic systems and (ii) Rule 54. In our calculations we will only work with the rescaled functions in Eq.~\eqref{eq:rescaledfuncts} and, to lighten the notation, we suppress the superscript $\sim$.

\subsection{Solution for Free Fermions}

Let us begin from the simplest case, i.e., free fermions. In this case Eqs.~\eqref{eq:SPrescaled} and \eqref{eq:Frescaled} simplify to 
\beq
\begin{split}
&\partial_t H^\q(x,t)+v_\q  \partial_x H^\q(x,t)=0\\
&\partial_t n_\q(x,t)+v_\q \partial_{x} n_\q (x,t)=0\\
&\ln\left[\frac{1-n^\q(x,0)}{n^\q(x,0)}\right] =\mu^\q-\frac{1}{2}\left[ ( H^\q(x,0)+ H^{-\q}x,0) \right]\\
&H^\q(x,\zeta)=2\lambda q_i^{\q} \,\text{B}_{1/2}(x)\,.
\label{SP_QQ-FCS_FF}
\end{split}
\eeq
and 
\beq
\begin{split}
f(\lambda,\mu, \zeta) &=-\frac{1}{2} \left\{\int \dd x \,(\textgoth{f}_\text{QQ}({\beta}(x,0))-\textgoth{f}_\text{QQ}(\mu))\!-\!\! \int \frac{\dd x}{2\pi}\, (\beta^\q(x,0)-\mu^\q)\!\cdot\! n_\q (x,0)\!-\!\! \int  \frac{\dd x}{2\pi}\, 2\lambda q_i^\q \!\cdot\! n_\q (x,\zeta)\,\text{B}_{1/2}(x)\right\}\,. 
\end{split}
\label{NonEqFCSWithSpSolFF}
\eeq
Here $v_\q= \varep'(\q)$ denotes the bare velocity and we used 
\be
\mathtt{A}={\rm diag}\{v_\q\}, \qquad  \beta^\q(x,t)=\ln\frac{1-n^\q(x,t)}{n^\q(x,t)}, \qquad \rho_\q (x,t) = \frac{1}{2\pi} {n_\q (x,t)}\,.
\ee
Note that in this case we also have a simple relation between $\mu^\q$ and the K-matrix of the initial state (cf.\ Sec.~\ref{IntegrableQuench}), but we will leave $\mu^\q$ implicit in the rest of the derivation. 

Since Eqs.~\eqref{SP_QQ-FCS_FF} do not involve non-linear dressings, they can be directly solved to give
\beq
\begin{split}
H^\q(x,t)&=H^\q(x-v_\q t,0)=2\lambda q^\q_i 
\,\text{B}_{1/2}(x+ v_\q (\zeta-t))\,,\\
\beta^\q(x,t)&=\beta^\q(x-v_\q t,0)=\mu^\q-\lambda q_i^\q \left(\,\text{B}_{1/2}(x+v_\q (\zeta-t))+\,\text{B}_{1/2}(x-v_{\q}(\zeta+t))  \right),\\
n_\q(x,t)&=n_\q(x-v_\q t ,0)=\left(1+e^{\mu^\q-\lambda q_i^\q \left(\,\text{B}_{1/2}(x+v_\q (\zeta-t))+\,\text{B}_{1/2}(x - v_\q (\zeta+t))  \right)}\right)^{-1},\\
\end{split}
\label{eq:explicitSolutionsFF}
\eeq
where we assumed that $q_i^\q$ and the dispersion relation $\varep_\q$ are symmetric functions of the rapidity. Then, noting that 
\begin{align}
\int \dd x\, (\beta^\q(x,0)-\mu^\q)\cdot n_\q (x,0) %& =  -\lambda \int\!\! \dd x \!\!\int \!\!\dd \q \, q_i(\q) \left(\,\text{B}_{1/2}(x+v(\q) \zeta)+\,\text{B}_{1/2}(x-v({\q}) \zeta)  \right) n(\q,x,0)\\
%& =  -\lambda \int\!\! \dd x \!\!\int \!\!\dd \q \, q_i(\q)  \text{B}_{1/2}(x) n(\q,x-v(\q) \zeta,0) -\lambda \int\!\! \dd x \!\!\int \!\!\dd \q \, q_i(\q)  \text{B}_{1/2}(x) n(\q,x+v(\q) \zeta,0) \\ 
& =  -\lambda \int\!\! \dd x\, q_i^\q  n_\q(x, \zeta)  \text{B}_{1/2}(x)  -\lambda \int\!\! \dd x \, q_i^\q   n_\q(x,-\zeta) \text{B}_{1/2}(x) \notag \\ 
& =  -2 \lambda \int\!\! \dd x\, q_i^\q  n_\q(x, \zeta)  \text{B}_{1/2}(x),
\end{align}
where in the first step we used Eqs.~\eqref{eq:explicitSolutionsFF} and in the second the evenness of $q_i^\q$ and $\varep_\q$, we see that second and third terms in Eq.~\eqref{NonEqFCSWithSpSolFF} cancel yielding  
\beq
\begin{split}
f(\lambda,\mu, \zeta) =&-\frac{1}{2} \int \dd x \,\textgoth{f}_\text{QQ}({\beta}(x,0))-\textgoth{f}_\text{QQ}(\mu)\\
=&-\frac{1}{2}\int  \frac{\dd \q}{2 \pi}  \left[\textgoth{f}^\q_\text{QQ}(\mu^\q-2\lambda q_i^\q)-\textgoth{f}^\q_\text{QQ}(\mu^\q)\right]\\
&+\frac{1}{2}\int \frac{\dd \q}{2 \pi} \text{min}(1,2|\varep'(\q)|\zeta)  \left[\textgoth{f}^\q_\text{QQ}(\mu^\q-2\lambda q_i^\q)+\textgoth{f}^\q_\text{QQ}(\mu^\q)-2\textgoth{f}^\q_\text{QQ}(\mu^\q-\lambda q_i^\q)\right], 
\end{split}
\eeq
which is the known exact result for free fermions~\cite{bertini2024dynamics}.

\subsection{Solution for Rule 54}

Let us now consider Rule 54. In this case we consider quenches from the initial states in Eq.~\eqref{eq:R54QuenchState}: a family of integrable quenches for which $\mu^+=\mu^-=\mu$, i.e., the chemical potentials associated with the left and right-moving particles coincide. Our focus is on the fluctuations of the total particle number. The saddle point equations specialised to this case read as 
\beq
\begin{split}
&\partial_t n_\nu(x,t)+v^\text{eff}_\nu [n_{-\nu}(x,t)] \partial_{x} n_\nu (x,t)=0\\
&\partial_t H^\nu(x,t)+ \mathtt{A}[\underline{n}(x,t)]_{\nu'}^{\,\,\,\nu} \partial_x H^{\nu'}(x,t)=0\\
&\beta^\nu[\underline{n}(x,0)]=\mu^\nu-\frac{1}{2} \left[ H^{\nu}(x,0)+ H^{-\nu}(x,0) \right]\\
&H^{\nu}(x,\zeta)=2\lambda \text{B}_{1/2}(x)\,,
\label{SP_QQ-FCS_WithH-R54}
\end{split}
\eeq
where $v^\text{eff}_\nu[n]=v^{\text{eff},\nu}[n]$.

The above equations are non-linear and far more complicated than those of the free fermionic case. To solve them, we assume that the solution takes a piece-wise constant form and verify our ansatz at the end of the calculation. Since $H^\nu$ satisfies the same linear equations as $\beta^\nu$ (up to boundary conditions), i.e., assuming a known solution for the chemical potential we can characterise the jumps of the $H$ functions over the the discontinuous regions. We assume the jumps of $\beta^\pm$ and $H^\pm$ across regions are proportional, leading to consistent equations once the geometry is specified. 

Below we present an explicit solution of Eqs.~\eqref{SP_QQ-FCS_WithH-R54} for a certain family of piece-wise constant initial conditions. Namely, we assume that at $t=0$, the initial filling functions are such that
\beq
n_+(x,0)=n_-(x,0)=
\begin{cases}
&n_0 \quad \text{if} \quad x<-a\\
&n_1 \quad \text{if} \quad -a\leq x<-b\\
&n_2 \quad \text{if} \quad -b\leq x \leq b\\
&n_1 \quad \text{if} \quad b < x\leq a\\
&n_0 \quad \text{if} \quad a < x\\
\end{cases}.
\label{FillingFunctiont0}
\eeq
with some $a,b>0$ positive numbers that are proportional to $X$. The functions $n_1$ and $n_2$ are unknown at this stage and we parametrise them as
\beq
n_1=\frac{1}{1+e^{\mu-\chi}}\,,\quad n_2=\frac{1}{1+e^{\mu-\upsilon}}\,,
\label{FillingFunctionParametrisation}
\eeq
and $\chi$ and $\upsilon$ are to be determined. To proceed, it is convenient to consider different time regimes, i.e.\ values of $\zeta$, separately.

\subsubsection{Solution for small $\zeta$}
\label{m=2Sol}

Let us begin by considering the case of small $\zeta$, which, as we shall see, includes the range $\zeta\leq 1/4$ where the FCS of Rule 54 is known exactly~\cite{bertini2024dynamics, klobas2024non}. Importantly, we will show that our results recover the latter. 

\begin{figure}
    \centering
    \vspace*{-2cm}
    \includegraphics[angle=-90, width=0.85\linewidth]{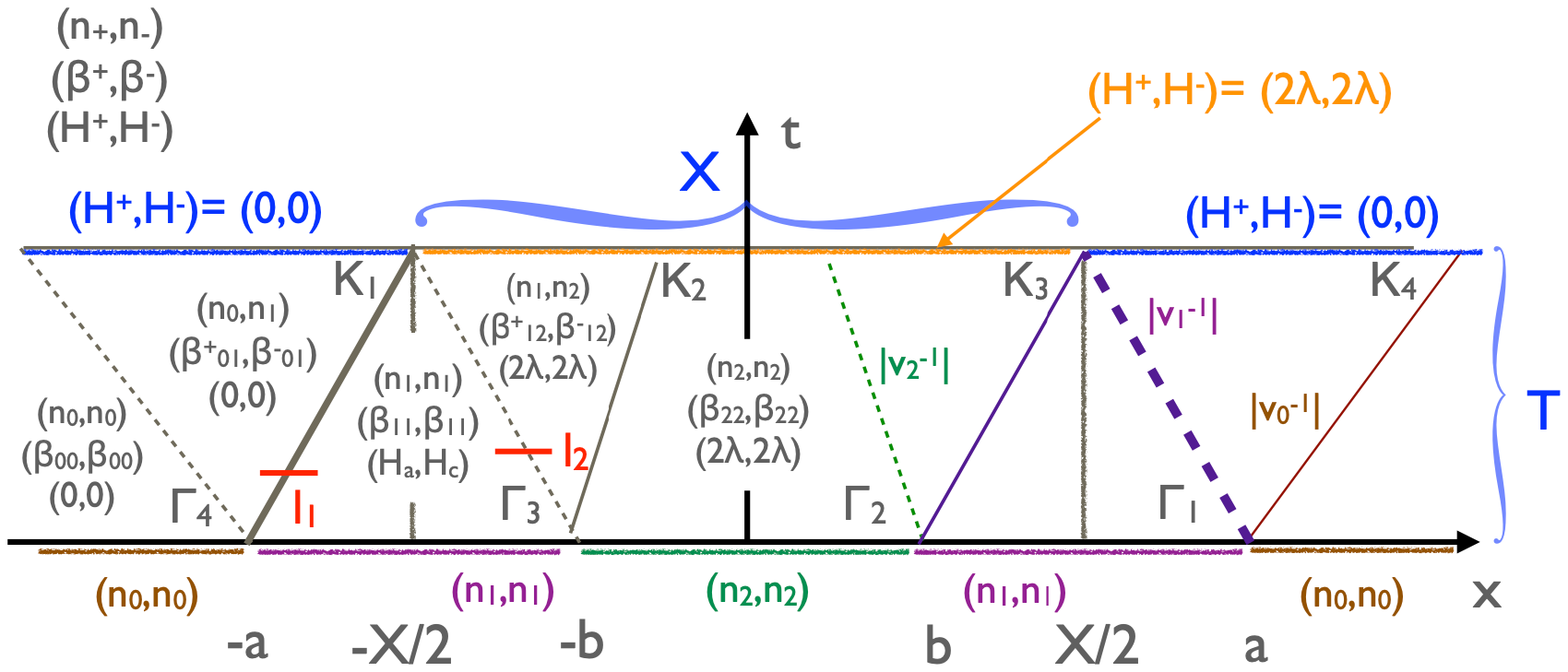}
    \vspace*{-2cm}
    \caption{Illustration of the consistent solution with 3-plateau piece-wise constant initial filling functions in the small-$\zeta$ ($T/X$) regime. The dashed lines correspond to characteristic curves that describe the propagation of the contact discontinuities to the left direction ($K_1$, $K_2$, $K_3$, and $K_4$). The solid lines correspond to characteristic curves that describe the propagation of the contact discontinuities to the right direction ($\Gamma_1$, $\Gamma_2$, $\Gamma_3$, and $\Gamma_4$). The slopes of these curves equal the inverse (effective) velocities; their absolute values are the same for lines with the same colour. In the left half, within each polygon, we showed the values of the various hydrodynamic fields in parentheses $(n_+,n_-), (\beta^+,\beta^-)$, and $(H^+, H^-)$; and for $\beta$, the lower indices refer to the right- and left-moving filling functions. To obtain these fields on the right side, the $+$ and $-$ components need to be swapped. We also show the initial condition for the filling function and the terminal condition for the $H$-field.}
    \label{fig:R54PDESolSmallT1}
\end{figure}

For sufficiently small $\zeta$, the evolution of the hydrodynamic fields $n_+(x,t)$ and $n_-(x,t)$ from the initial condition in Eq.~\eqref{FillingFunctiont0} follows the pattern illustrated in Fig.~\ref{fig:R54PDESolSmallT1}. There we introduced eight curves denoting the evolution of the boundaries of the piece-wise constant regions of $n_+$ and $n_-$. Namely, $K_1,\ldots,K_4$ (solid) track the successive jumps in $n_+$ and $\Gamma_1,\ldots,\Gamma_4$ (dashed) track the corresponding ones in $n_-$. Within each polygon bounded by these curves the fields are constant: their values $(n_+, n_-)$ are reported in Fig.~\ref{fig:R54PDESolSmallT1}. The hydrodynamic fields $(n_+, n_-)$ are Riemann invariants, hence constant along characteristics determined by the geometry. The slopes of $K$/$\Gamma$ are the inverse effective velocities and, by linear degeneracy, each one depends only on $n_{\mp}$. For $\lambda>0$ we have $v_2<v_1<v_0$. The geometry described in Fig.~\ref{fig:R54PDESolSmallT1} implies that the characteristic lines $K_2$ and $\Gamma_2$ cross the $t=T$ line respectively at positions $-{x}_*$ and ${x}_*$, with $x_*$ arbitrary positive number. This means that the small $\zeta$ regime is defined by 
\beq
\zeta \le \zeta_1=\frac{1}{2(v^\text{eff}(n_1)+v^\text{eff}(n_2))}\le \frac{3}{8}\,,
\label{Zetam2}
\eeq
which, as anticipated, is larger than the regime of validity of the exact result~\cite{bertini2024dynamics, klobas2024non}. 

We now proceed with the solution. We begin by recalling that $\beta_\pm$ are functionals of $n_\pm$ and, importantly, they can be discontinuous only where $n_\pm$ are. This means that the values of $(\beta_+,\beta_-)$ follow the pattern depicted in Fig.~\ref{fig:R54PDESolSmallT1}, where in each region $\beta_\pm(n_+,n_-)$ are smooth functions (not necessarily constant) and we used that 
\be
\label{eq:nodressing}
\beta_+(n,n) = \beta_-(n,n) \equiv \beta(n,n)
\ee
as the interaction terms in Eq.~\eqref{eq:betavsn} cancel for equal fillings in Rule 54. 

Next, we recall that $H_\nu$ and $\beta_\nu$ fulfil the same equation. Moreover, the terminal conditions (at $t=\zeta$) for $H_\nu$ in Eqs.~\eqref{SP_QQ-FCS_WithH-R54} have discontinuities only where the profiles of $\beta_\nu$ for $t=\zeta$ have discontinuities. This means that one can propagate $H_\nu$ backwards in time obtaining the profiles reported in Fig.~\ref{fig:R54PDESolSmallT1}. In particular, due to the piecewise constant form of the terminal condition we have that $(H_+,H_-)$ is fully specified in all regions apart from those comprised between $K_1$ and $\Gamma_3$ and between $K_3$ and $\Gamma_1$, where, by symmetry, it takes the same value. We denote the (unknown) values of $H_+$ and $H_-$ in those regions by $H_a$ and $H_c$ respectively. Imposing now the initial condition for $\beta_\nu$ (cf.\ Eqs.~\eqref{SP_QQ-FCS_WithH-R54}) we then have 
\be
a= \frac{1}{2}+\zeta |v^\text{eff}(n_1)|, \qquad b=\frac{1}{2}-\zeta |v^\text{eff}(n_1)|,
\ee
and
\beq
\chi=(H_a+H_c)/2\,,\qquad\qquad \upsilon=2\lambda\,.
\label{eq:initialbeta}
\eeq
Now we use our assumption on the proportionality between the jumps of $H$ and $\beta$ to obtain 2 equations describing the jumps $l_1$ and $l_2$ in Fig.~\ref{fig:R54PDESolSmallT1}, i.e., 
\beq
\begin{split}
&\begin{pmatrix}
H_a\\
H_c
\end{pmatrix}=\alpha \begin{pmatrix}
\beta^+_{11}-\beta^+_{01}\\
\beta^-_{11}-\beta^-_{01}
\end{pmatrix}\,,\quad \begin{pmatrix}
2\lambda-H_a\\
2\lambda-H_c
\end{pmatrix}=
\gamma \begin{pmatrix}
\beta^+_{12}-\beta^+_{11}\\
\beta^-_{12}-\beta^-_{11}
\end{pmatrix}
\,.
%\qquad h_b=2\lambda\,,\quad \text{or}\quad  n_1=n_2\,.
\end{split}
\label{JumpConditionsGeom1}
\eeq
Solving these equations we can express $H_a, H_c$ and $\alpha, \gamma$ in terms of the $\beta$ fields or, by using Eq.~\eqref{eq:betavsn}, of $n_0$, $n_1$, and $n_2$. Recalling the parametrisation in Eq.~\eqref{FillingFunctionParametrisation} we turn Eq.~\eqref{eq:initialbeta} into a consistency condition for the variable $\chi$, i.e, we can write
\beq
\chi=\frac{H_a(n_0, n_1, n_2)+H_c(n_0, n_1, n_2)}{2}=\frac{H_a(\mu,\lambda, \chi)+H_c(\mu,\lambda, \chi)}{2}. 
\label{SmallTFinalSelfConsEqs}
\eeq
This equation is reported explicitly in Appendix \ref{AppSmallT}, where we show that pruning away unphysical solutions (giving, e.g., complex values of the $H$ fields, or zero or infinite proportionality constants), the equation can be reduced to  
\beq
\left(n_0+\frac{1-n_0}{z}\right)^2=\frac{e^\lambda(1-n_0+e^{2\lambda}n_0)}{z^3}\,,
\label{eq:simplecubic}
\eeq
where we set $z=e^\chi$ and used $n_0=(1+e^\mu)^{-1}$. This equation has a unique real solution, which exists for arbitrary real $\mu$ and $\lambda$, and can be expressed as 
\be
\chi^*=\ln\left(\frac{\sqrt[3]{\sqrt{\kappa ^2-4 e^{6 \mu }}+\kappa }}{3 \sqrt[3]{2}}+\frac{\sqrt[3]{2} e^{2 \mu }}{3 \sqrt[3]{\sqrt{\kappa ^2-4 e^{6 \mu }}+\kappa }}-\frac{2 e^{\mu }}{3}\right),
\label{Solution}
\ee
with 
\be
\kappa=27 e^{\lambda +\mu }+27 e^{3 \lambda +\mu }+27 e^{\lambda +2 \mu }+27 e^{3 \lambda }+2 e^{3 \mu }. 
\ee
In Appendix \ref{AppSmallT} we demonstrate that the above solution is unique and, therefore, it is the only physical solution of Eq.~\eqref{SmallTFinalSelfConsEqs}.

Specialising Eq.~\eqref{eq:Frescaled} to the case of Rule 54 and using the solutions we found for $n_\nu$, $\beta_\nu$, and $H_\nu$ we have 
\begin{align}
f(\lambda,\mu,\zeta)=& -4\zeta |v^\text{eff}(n_1)|\left(\ln(1-n_1)+\ln(1+e^{-\mu})\right)-\left(1-2\zeta (|v^\text{eff}(n_1)|)\right)\left(\ln(1-n_2)+\ln(1+e^{-\mu})\right)\notag\\
&+ 4\zeta |v^\text{eff}(n_1)| \left(\ln\left(\frac{1-n_1}{n_1}\right)-\mu\right) n_1+\left(1-2\zeta |v^\text{eff}(n_1)|\right) \left(\ln\left(\frac{1-n_2}{n_2}\right)-\mu\right)n_2\\
&+2\lambda \zeta (|v^\text{eff}(n_1)|+|v^\text{eff}(n_2)|)\left[\frac{1+2n_1}{1+n_1+n_2}n_2+\frac{1+2n_2}{1+n_1+n_2}n_1\right]+2\lambda \left(1-2 \zeta (|v^\text{eff}(n_1)|+|v^\text{eff}(n_2)|)\right) n_2\,,\notag
\end{align}
which can be simplified to 
\be
\frac{f(\lambda,\mu,\zeta)-f(\lambda,\mu,0)}{\zeta} =\frac{4}{1+2n_1}\left\{2\ln\left(n_0 e^{\chi^*}+1-n_0\right) -\ln (1-n_0+e^{2\lambda}n_0) + 2 n_1 (\lambda-\chi)\right\} =4(\lambda-\chi^*),
\label{Slope_Gen}
\ee
where in the last step we used Eq.~\eqref{eq:simplecubic}.\\

\noindent \underline{Comparison with the exact solution}. For $\zeta \leq 1/4$ the non-equilibrium FCS for the R54 model has been computed exactly in Ref.~\cite{bertini2024dynamics, klobas2024non}. In our conventions, the solution is expressed as 
\beq
\frac{f_\text{exact}(\lambda,\mu,\zeta)-f(\lambda,\mu,0)}{\zeta} = 4 \textgoth{m},
\eeq
where $\textgoth{m}$ is defined as
\beq
\textgoth{m}=\ln(w^*)-\ln{(1-n_0+e^{2\lambda}n_0)},
\label{BKSlope}
\eeq
and $w^*$ is the largest solution of the cubic equation
\beq
w^3=((1-n_0)w+n_0 e^{\lambda}(1-n_0+e^{2\lambda}n_0))^2\,. 
\label{BKCubic}
\eeq
This equation reduces to Eq.~\eqref{eq:simplecubic} if one sets 
\be
w = \frac{e^{\lambda}(1-n_0+e^{2\lambda}n_0)}{z},
\ee
which immediately establishes the same correspondence among their unique real solutions, i.e.
\be
w^* =  {e^{\lambda}(1-n_0+e^{2\lambda}n_0)} e^{- \chi^*}\,. 
\ee
Substituting this into Eq.~\eqref{BKSlope} we do recover our result in Eq.~\eqref{Slope_Gen}.

\subsubsection{Solution for intermediate $\zeta$}
\label{SmallTExtending}

The piece-wise constant initial condition in Eq.~\eqref{FillingFunctiont0}, with $\chi$ and $\upsilon$ given by Eqs.~\eqref{eq:initialbeta} and \eqref{Solution}, continues to be a solution of Eqs.~\eqref{SP_QQ-FCS_WithH-R54} for 
\beq
\zeta_1\leq \zeta \leq \zeta_2=\frac{1}{2v^\text{eff}(n_1)}, 
\eeq
with some $a>b>0$ positive numbers that are proportional to $X$. 

\begin{figure}
    \centering
    \vspace*{-2cm}
    \includegraphics[angle=-90, width=0.85\linewidth]{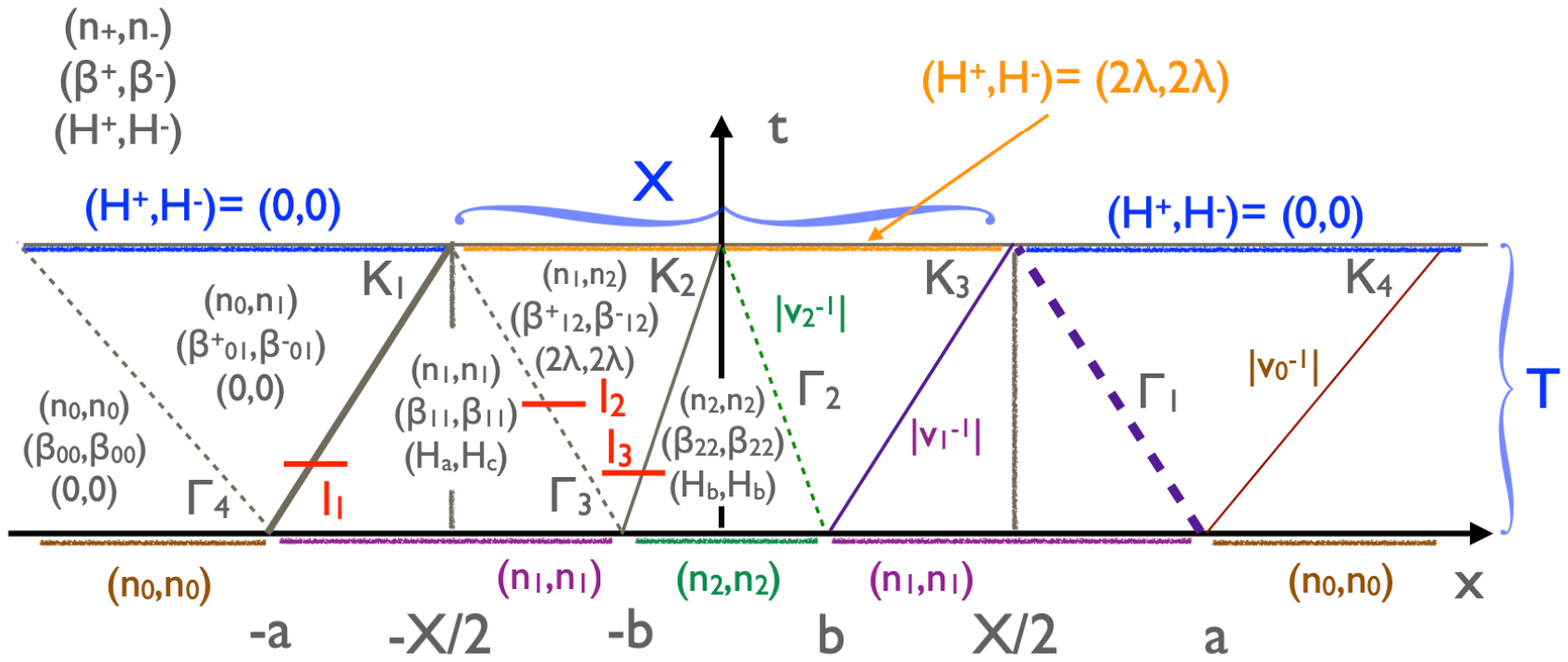}
    \vspace*{-2cm}
    \caption{Illustration of the consistent solution with a 3-plateau piece-wise constant initial filling functions when $T/X=\zeta=\zeta_1$. The dashed lines correspond to characteristic curves that describe the propagation of the contact discontinuities to the left direction ($K_1$, $K_2$, $K_3$, and $K_4$). The solid lines correspond to characteristic curves that describe the propagation of the contact discontinuities to the right direction ($\Gamma_1$, $\Gamma_2$, $\Gamma_3$, and $\Gamma_4$).  The slopes of these curves equal the inverse (effective) velocities; their absolute values are the same for lines with the same colour. In the left half, within each polygon, we showed the values of the various hydrodynamic fields in parentheses $(n_+,n_-), (\beta^+,\beta^-)$, and $(H^+, H^-)$; and for $\beta$, the lower indices refer to the right- and left-moving filling functions. To obtain these fields on the right side, the $+$ and $-$ components need to be swapped. We also show the initial condition for the filling function and the terminal condition for the $H$-field.}
    \label{fig:R54PDESolSmallT2}
\end{figure}

\begin{figure}
    \centering
    \vspace*{-2cm}
    \includegraphics[angle=-90, width=0.85\linewidth]{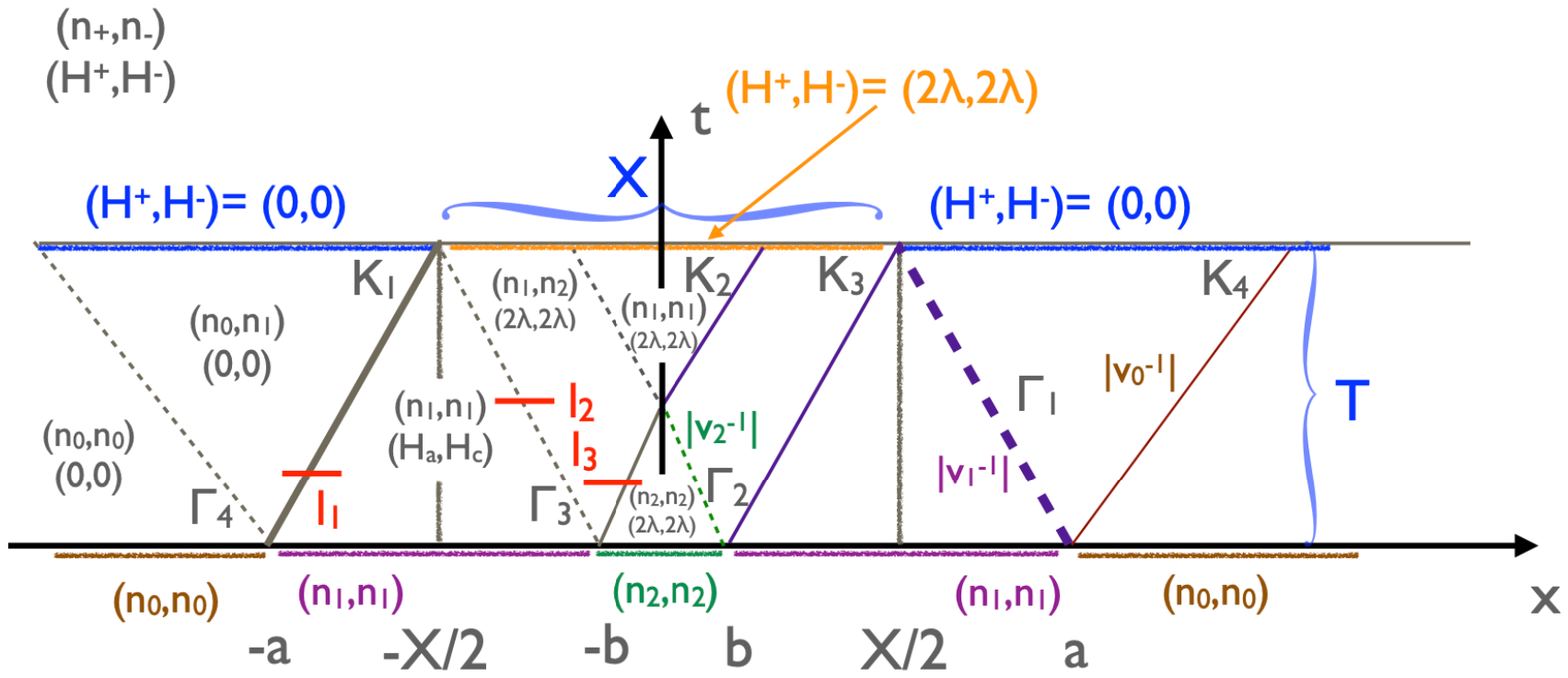}
    \vspace*{-2cm}
    \caption{Illustration of the consistent solution with a 3-plateau piece-wise constant initial filling functions when $\zeta_2>\zeta=T/X>\zeta_1$. The dashed lines correspond to characteristic curves that describe the propagation of the contact discontinuities to the left direction ($K_1$, $K_2$, $K_3$, and $K_4$). The solid lines correspond to characteristic curves that describe the propagation of the contact discontinuities to the right direction ($\Gamma_1$, $\Gamma_2$, $\Gamma_3$, and $\Gamma_4$).  The slopes of these curves equal the inverse (effective) velocities; their absolute values are the same for lines with the same colour. In the left half, within each polygon, we showed the values of the various hydrodynamic fields in parentheses $(n_+,n_-)$ and $(\beta^+,\beta^-)$; and for $\beta$, the lower indices refer to the right- and left-moving filling functions. To obtain these fields on the right side, the $+$ and $-$ components need to be swapped. We also show the initial condition for the filling function and the terminal condition for the $H$-field.}
    \label{fig:R54PDESolSmallT3}
\end{figure}

To see this, consider first the case of $\zeta=\zeta_1$, which is illustrated in Fig.~\ref{fig:R54PDESolSmallT2}. The only difference compared to the previous case (cf.~Fig.~\ref{fig:R54PDESolSmallT1}), is that the value of the $H$ fields in the central triangular region is not immediately obvious (we denote it as $H_b$). To find it, we write jump conditions for the jumps $l_1$, $l_2$, and $l_3$ marked in red in the figure
\beq
\begin{split}
&\begin{pmatrix}
H_a\\
H_c
\end{pmatrix}=\alpha \begin{pmatrix}
\beta^+_{11}-\beta^+_{01}\\
\beta^-_{11}-\beta^-_{01}
\end{pmatrix}\,,\quad \begin{pmatrix}
2\lambda-H_a\\
2\lambda-H_c
\end{pmatrix}=
\gamma \begin{pmatrix}
\beta^+_{12}-\beta^+_{11}\\
\beta^-_{12}-\beta^-_{11}
\end{pmatrix}
\,,\quad \begin{pmatrix}
H_b-2\lambda\\
H_b-2\lambda
\end{pmatrix}=
\delta \begin{pmatrix}
\beta_{22}-\beta^+_{12}\\
\beta_{22}-\beta^-_{12}
\end{pmatrix}, 
\end{split}
\eeq
where we again used Eq.~\eqref{eq:nodressing}.
Note that $n_1$ must be different from $n_2$ for $\lambda \neq 0$. To see this, let us momentarily assume the opposite, namely that $n_1=n_2\neq n_0$. In this case the the values of $H$ in the central region spanned by thy trapezium with lower vertices as $-a$ and $a$, and upper vertices at $-X/2$ and $X/2$ in Fig. \ref{fig:R54PDESolSmallT2} are identically $2 \lambda$ due to the terminal condition at $t=T$. This means that there is one one non-trivial jump at the segment $l_1$. The jump in $H$ is non-zero and is proportional to $(1,1)$, however, the non-zero jump in $\beta$ cannot be proportional to $(1,1)$ if $n_0\neq n_1$ due to TBA equations (different $n^\pm\rightarrow$  different $\beta^\pm$). One can similarly rule out the possibility of $n_0=n_1 \neq n_2$.
Since, $n_1$ must be different from $n_2$, we have that $\beta^+_{12} \neq \beta^-_{12}$ and, therefore, the third equation can only be satisfied if $h_b=2\lambda$ and $\delta=0$. This means that the equations that determine the unknown filling functions remain exactly the same as in the $\zeta<\zeta_1$ region. This observation is important, as now we can further shrink the extent of the region where the initial filling function equals $n_2$, as illustrated by Fig. \ref{fig:R54PDESolSmallT3}. In this case, the types of boundaries coincide with those in Fig. \ref{fig:R54PDESolSmallT2}. Therefore, the solutions for the $n^\pm$ fields in the $\zeta_2 \ge \zeta > \zeta_1$ are the same as in the $\zeta_1 \ge \zeta$ regime.

An important recognition is that this solution can only be valid up until the two equilateral triangles with slopes $|v_1^{-1}|$ touch each other, which results in the upper limit for $\zeta$
\beq
\zeta_2=\frac{1}{2v^\text{eff}(n_1)}\,.
\eeq
It is easy to check that the FCS in this $\zeta$ regime equals Eqs.~\eqref{Slope_Gen}. 

\subsubsection{Solution for large $\zeta$}
\begin{figure}
    \vspace*{-2cm}
    \centering
\includegraphics[angle=-90,width=0.85\linewidth]{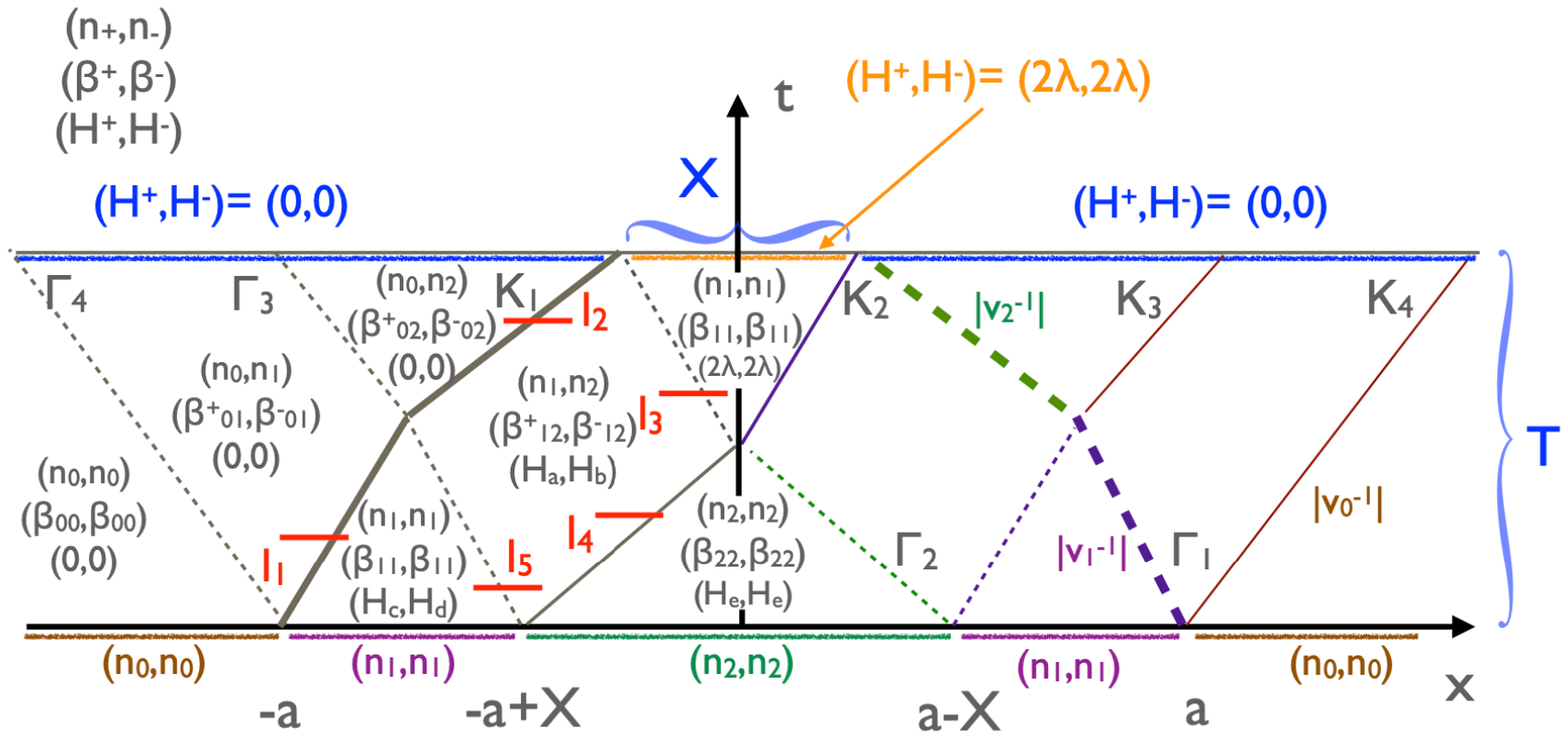}
    \vspace*{-2cm}
\caption{Illustration of the consistent solution with 3-plateau piece-wise constant initial filling functions in the large-$\zeta$ ($T/X$) regime. The dashed lines correspond to characteristic curves that describe the propagation of the contact discontinuities to the left direction ($K_1$, $K_2$, $K_3$, and $K_4$). The solid lines correspond to characteristic curves that describe the propagation of the contact discontinuities to the right direction ($\Gamma_1$, $\Gamma_2$, $\Gamma_3$, and $\Gamma_4$).  The slopes of these curves equal the inverse (effective) velocities; their absolute values are the same for lines with the same colour. In the left half, within each polygon, we showed the values of the various hydrodynamic fields in parentheses $(n_+,n_-), (\beta^+,\beta^-)$, and $(H^+, H^-)$; and for $\beta$, the lower indices refer to the right- and left-moving filling functions. To obtain these fields on the right side, the $+$ and $-$ components need to be swapped. We also show the initial condition for the filling function and the terminal condition for the $H$-field.}
\label{fig:PDESolGGE}
\end{figure}

Another regime that can be characterised exactly is that of large $\zeta$. Indeed, using the exact characterisation of the late-time regime provided in Refs.~\cite{klobas2021exact, klobas2021exact2}, one can directly show that if $\zeta \geq (2\, \text{inf} |v^\text{eff}|)^{-1}= 0.75$ one finds
\be
f(\lambda,\mu,\zeta)=-\left( \,\textgoth{f}_\text{QQ}(\mu-\lambda)-\textgoth{f}_\text{QQ}(\mu)\right)=f_\text{GGE}(\lambda,\mu),
\label{eq:GGEFCS}
\ee
which describes the FCS in the GGE reached after a quench from the initial state in Eq.~\eqref{eq:R54QuenchState}. This result can again be recovered using our Eqs.~\eqref{SP_QQ-FCS_WithH-R54} following steps very similar to the ones illustrated above. In this case the evolution of the hydrodynamic fields $n_+(x,t)$ and $n_-(x,t)$ from the initial condition in Eq.~\eqref{FillingFunctiont0} follows the pattern illustrated in Fig.~\ref{fig:PDESolGGE}. The latter holds as long as $b=v^\text{eff}(n_2))(\zeta-1/(2 v^\text{eff}(n_1))))\geq0$, that is,
\beq
\zeta \geq (2 v^\text{eff}(n_1))^{-1}\,. 
\label{R54zetaGGE}
\eeq
Using again our proportionality assumption, we see that the jumps $l_1, \ldots, l_5$ in Fig.~\ref{fig:PDESolGGE} are defined by 
\beq
\begin{split}
&\begin{pmatrix}
\beta^+_{11}-\beta^+_{10}\\
\beta^-_{11}-\beta^-_{10}
\end{pmatrix}
=\alpha \begin{pmatrix}
H_c\\
H_d
\end{pmatrix}
\,, \quad \begin{pmatrix}
\beta^+_{21}-\beta^+_{20}\\
\beta^-_{21}-\beta^-_{20}
\end{pmatrix}=\gamma \begin{pmatrix}
H_a\\
H_b
\end{pmatrix}\,, \quad
\begin{pmatrix}
\beta^+_{11}-\beta^+_{21}\\
\beta^-_{11}-\beta^-_{21}
\end{pmatrix}=\delta \begin{pmatrix}
2\lambda-H_a\\
2\lambda-H_b
\end{pmatrix}\\
&\begin{pmatrix}
\beta^+_{22}-\beta^+_{21}\\
\beta^-_{22}-\beta^-_{21}
\end{pmatrix}=\epsilon \begin{pmatrix}
H_e-H_a\\
H_e-h_b
\end{pmatrix}\,,\quad \begin{pmatrix}
\beta^+_{21}-\beta^+_{11}\\
\beta^-_{21}-\beta^-_{11}
\end{pmatrix}=\phi \begin{pmatrix}
H_a-H_c\\
H_b-H_d
\end{pmatrix}.
\end{split}
\label{JumpConditionsGeom4}
\eeq
As before, we can combine these equations with Eq.~\eqref{eq:betavsn} to express $H_a, H_b,...,H_e, \alpha, \gamma,...,\phi$ in terms of $n_0,n_1$ and $n_2$. Then, considering the linear combinations 
\beq
\begin{split}
\chi&=\frac{H_c(n_0, n_1, n_2)+H_d(n_0, n_1, n_2)}{2}=\frac{H_c(\mu,\chi, \upsilon)+H_d(\mu,\chi, \upsilon)}{2}\\
\upsilon&=H_e(n_0, n_1, n_2)=H_e(\mu,\chi, \upsilon)\\
\label{LargeTFinalSelfConsEqs}
\end{split}
\eeq
we obtain closed consistency equations for the unknown variables $\chi$ and $\upsilon$. The explicit form of these equations, and their analysis, are presented in Appendix \ref{AppLargeT}, while here we only discuss the results, that in this case are indeed very simple: the only physical solution of Eqs.~\eqref{LargeTFinalSelfConsEqs} reads as  
\beq
\chi=\lambda\,,\quad \upsilon=0\,. 
\eeq
Plugging it into Eq.~\eqref{FillingFunctiont0} and then into Eq.~\eqref{eq:Frescaled} specialised to Rule 54 we obtain Eq.~\eqref{eq:GGEFCS} for all $\zeta>(2v^\text{eff}(n_1|_{\chi=\lambda}))^{-1}\equiv (2 v_\lambda)^{-1}$. Since $(2 v_\lambda)^{-1} < (2\, \text{inf} |v^\text{eff}|)^{-1}= 0.75 $, we conclude that our BMFT treatment recovers. 

At this point, however, we note a possible source of inconsistency. Recalling that the $v_\lambda$ defined above does not coincide with the $v^\text{eff}(n_1)$ obtained in the previous subsections because in the latter $n_1$ is computed with $\chi=\chi^*$ (cf.~Eq.~\eqref{Solution}), we have that large and small $\zeta$ solutions can overlap. This inconsistency is resolved in the upcoming subsection.

\subsubsection{The complete solution for arbitrary $\zeta$ in R54}

\begin{figure}
    \centering
    \includegraphics[angle=-90,width=0.7\linewidth]{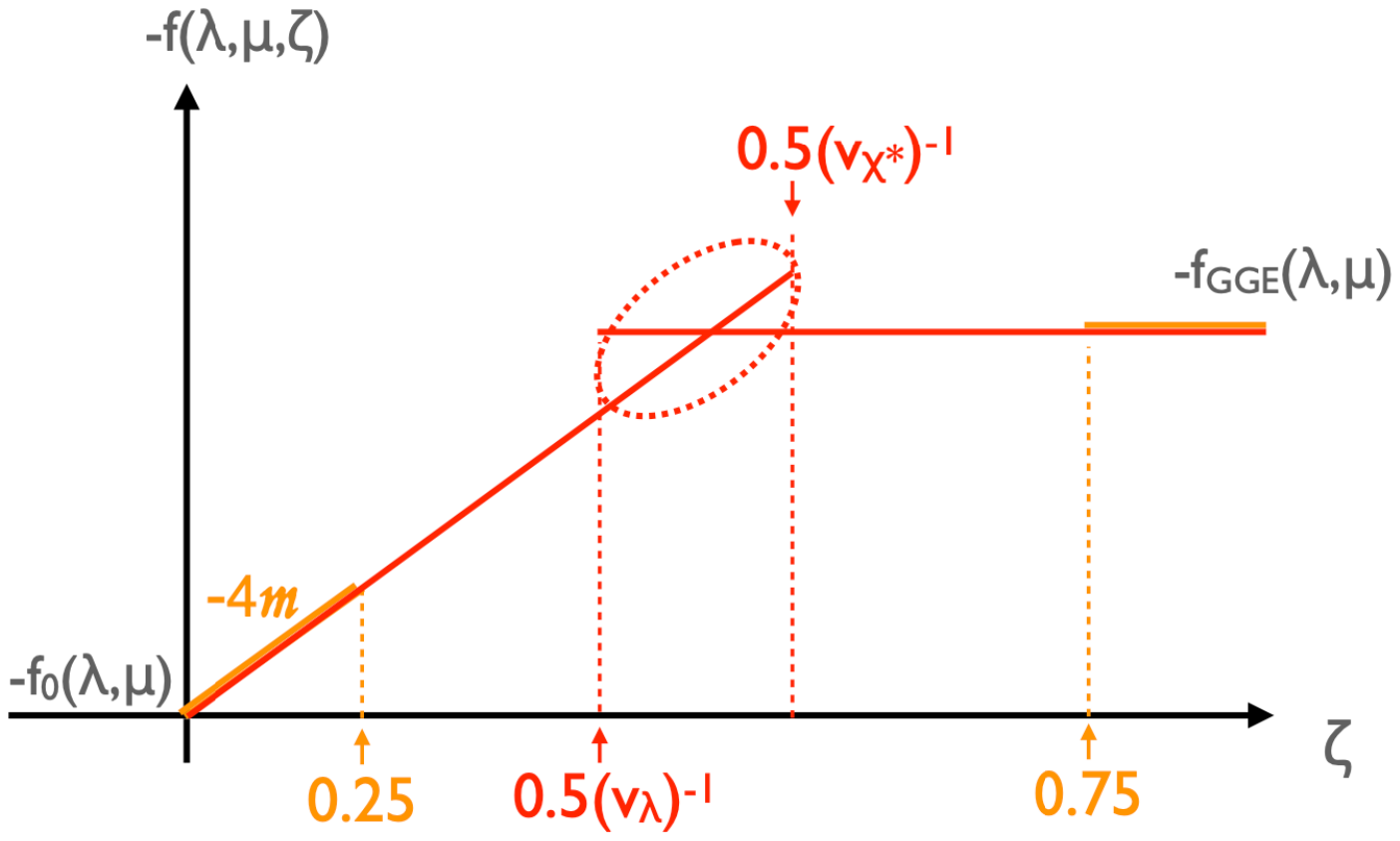}
    \vspace*{-1cm}
\caption{Illustration of two branches of analytic solutions for the time evolutions of the FCS as a function of $\zeta=T/X$ (red lines). The orange lines denote the exact microscopic predictions using space-time swap ideas.}
\label{fig:R54SolBranches}
\end{figure}

In this subsection we combine the results of the previous subsections to find a solution for the FCS valid for all $\zeta$s. Summing up our results so far, we have found that the FCS can be written as 
\beq
f(\lambda,\mu, \zeta)=
\begin{cases}
\frac{1}{2}f_\text{GGE}(2\lambda,\mu)+4 \textgoth{m} \zeta \,,\quad \text{if}\quad \zeta \leq  (2 v_{\chi^*})^{-1}\\
f_\text{GGE}(\lambda,\mu)\,,\qquad \qquad\quad \text{if}\quad \zeta \geq  (2 v_\lambda)^{-1}\\
\end{cases}
\label{R54SolWithBranches}
\eeq
where we set $v_{\chi^*}\equiv v^\text{eff}(n_1|_{\chi=\chi^*})$ and we recall that 
\beq
 f_{\mathrm{GGE}}(\lambda,\mu)=-2\log(1+e^{-\mu})+2\log\bigl(1+e^{-(\mu-\lambda)}\bigr).
 \label{GGEFCSForBranches}
\eeq
First we note that the branches above overlap, namely, as we show in Appendix \ref{ProofForBranches}, we have 
\beq
\begin{split}
\text{i)}&\qquad(v_{\chi^*})^{-1} \geq (v_\lambda)^{-1} \Leftrightarrow n_1 \geq n_\lambda\\
\text{ii)}&\qquad f_\text{GGE}(\lambda,\mu)-\frac{1}{2}f_\text{GGE}(2\lambda,\mu) \geq 2\frac{\textgoth{m}}{v^\text{eff}_{\chi^*}}. 
\end{split}
\eeq
Therefore, the situation is as the one illustrated in Fig.~\ref{fig:R54SolBranches}. In this case one can obtain a unique physical solution by demanding the FCS to be continuous with respect to $\zeta$. This gives
\beq
f(\lambda,\mu, \zeta)=
\begin{cases}
\frac{1}{2}f_\text{GGE}(2\lambda,\mu)+4 \textgoth{m} \zeta \,,\quad \text{if}\quad \zeta \leq  \zeta^*\\
f_\text{GGE}(\lambda,\mu)\,,\qquad \qquad\quad \text{if}\quad \zeta \geq  \zeta^*\\
\end{cases}
\,\, \text{with}\quad 
\zeta^*=\frac{f_\text{GGE}(\lambda, \mu)-1/2 f_\text{GGE}(2\lambda, \mu)}{4\textgoth{m}}\,,
\label{R54SolWithoutBranches}
\eeq
where $\zeta^*$ will be shown to be always positive. 

Since we are working in an asymptotic scaling limit (cf.~Eq.~\eqref{EulerCorrelators}), however, it is not obvious why we should require continuity and it would be desirable to find a stronger physical principle allowing us to choose a branch. Since the branches of Eq.~\eqref{R54SolWithBranches} coexist, it is tempting to resort to a form of least action principle, which is, however, not immediately justified. The main issue is that the sign of the FCS is important and the way it is defined (cf.~Eq.~\eqref{GGEFCSForBranches}) ensures its convexity with respect to $\lambda$, and hence the existence of a rate function associated with the rare but large fluctuation of the charge. From this view-point, merely flipping the sign of the entire functional, i.e.\ applying a ``greatest action'' principle, is not motivated. However, we can formulate a least-action-type argument as follows. We first recall the (general) functional relation between the FCS in the initial state and in the steady-state GGE
\beq
f_0(\lambda,{\mu})=\frac{1}{2}f_\text{GGE}(2\lambda,{\mu})\,,
\label{InitFCS-GGE}
\eeq
and that $f_\text{GGE}(\lambda,{\mu})=f_\text{GGE}({\mu})-f_\text{GGE}({\mu}-\lambda)$, which guarantees the convexity of $f_\text{GGE}(\lambda,{\mu})$ as well as the fact that $f_\text{GGE}(0,{\mu})=0$. Because of these properties and of Eq.~\eqref{InitFCS-GGE} it is immediate that
\beq
f_\text{GGE}(\lambda,{\mu})\leq \frac{1}{2}f_\text{GGE}(2\lambda,{\mu})=f_0(\lambda,{\mu})\,,
\eeq
i.e., for each value of the counting field, the initial FCS is larger than the FCS in the steady-state. Whereas the end-points of the dynamics do not constrain the monotonicity property of the time evolution itself, based on comparing the initial and steady-state fluctuations, we can claim that they overall decrease over the course of the dynamics. Therefore, we can demand the slowest possible decrement of the FCS as a function of $\zeta$, or equivalently, the slowest possible increment of $-f$. This removes the ambiguity and gives us Eq.~\eqref{R54SolWithoutBranches}.

\section{Conclusions}
\label{sec:conclusions}

In this work we extended the ballistic macroscopic fluctuation theory to integrable quantum quenches, i.e., to a genuinely nonequilibrium setting. The resulting construction provides a general framework for describing large-scale dynamics of observables possessing a nontrivial Euler-scale limit. As a non-trivial example, we considered the time-dependent FCS of a conserved charge in a macroscopic subsystem, which directly probes the evolving large-scale fluctuations described by the theory. While the subsequent Euler-scale evolution is still governed by the deterministic GHD equations, just as in ``conventional'' BMFT, the initial fluctuation measure is no longer the standard one associated with a local GGE. Instead, it is obtained from the Quench Action functional and constrained to root-density profiles satisfying the pairing condition
\(\rho_\q(x,0)=\rho_{-\q}(x,0)\). In this way, the Quench Action determines the statistics of the initial hydrodynamic fields, while GHD transports these fluctuations through space and time.

We formulated a BMFT--Quench Action construction as a path integral over fluctuating root-density profiles. The pair constraint and the GHD continuity equation were imposed by auxiliary fields, and the Euler-scale limit reduced the path integral to a saddle-point problem. The resulting equations consist of the GHD equation for the root densities, a conjugate hydrodynamic equation for the counting field \(H^\q\), and boundary conditions coupling the initial thermodynamic state to the observable inserted at the final time. This insertion corresponds to biasing the measure that described the relevant fluctuations. For the FCS of a charge contained in a single interval, the problem depends on the interval length \(X\) and the evolution time \(T\) only through the ballistic ratio
\(\zeta=T/X\) at the relevant scales.

We tested the resulting framework in two complementary settings: noninteracting free-fermion systems and the interacting integrable cellular automaton Rule~54. For free fermionic systems the saddle-point equations decouple into simple advection equations and can be solved directly by the method of characteristics. The resulting FCS agrees with the known microscopic expression. This provides a first check that the pair-constrained initial measure and its deterministic Euler-scale propagation correctly reproduce the dynamics of charge fluctuations.

We then applied the construction to Rule~54, where interactions make both the hydrodynamic evolution and the conjugate equation nonlinear. Exploiting the linear degeneracy of the Rule~54 hydrodynamics, we constructed piecewise-constant saddle-point solutions separated by propagating contact discontinuities. In the small-\(\zeta\) regime, the consistency conditions reduce to a cubic equation with a unique physical solution. The resulting FCS reproduces the exact microscopic result in its previously established regime of validity and extends the same branch to larger values of \(\zeta\). At sufficiently large \(\zeta\), a second solution gives the FCS of the stationary GGE, as expected once the charge fluctuations inside the interval have locally relaxed to their late-time form. Remarkably, both analytic solutions remain valid beyond the regimes in which they were originally derived. Although their extended domains do not coincide completely, they overlap over a finite space-time window, and we prove that the two branches always cross within this common domain. Continuity of the FCS, or, equivalently, selection of the slowest admissible decrease of its fluctuations, then uniquely determines how the branches must be joined. This yields a new exact Euler-scale solution for arbitrary $\zeta$, including the overlapping regime, that is compatible with the hydrodynamic equations and the required physical and analytical properties of the FCS.

The general BMFT-QA construction assumes that the conserved charge under consideration possesses a nontrivial extensive FCS in the initial state. As discussed in Appendix~\ref{app:overlap}, this condition is expected to hold generically for finite-range parity-even charges in dimer-type integrable initial states. The principal exceptions are charges for which the initial state is an exact eigenstate, such as symmetry charges fixed by the preparation protocol, as well as reflection-odd charges annihilating an integrable initial state. Such non-fluctuating directions lead to singular constraints on the root-density measure and require a separate treatment.

Several extensions appear natural. A particularly important direction is to seek further exact solutions of the saddle-point equations in interacting Bethe-ansatz-solvable models. Exact conjectures are already available for the initial slope of the FCS with respect to $\zeta$~\cite{travaglino2026space}, and recovering these results would provide a stringent test of the framework. The corresponding small-$\zeta$ regime appears especially promising, as it may be accessible through a controlled solution of the saddle-point equations without requiring their complete solution at arbitrary $\zeta$. The method is also immediately applicable to systems with a discrete set of hydrodynamic normal modes, among which integrable cellular automata provide particularly tractable examples. In these models, one may go beyond a single interval and study the joint FCS associated with several intervals in different relative positions and alignments, thereby accessing a richer class of space-time fluctuation geometries. A central difficulty in both analytical and numerical treatments is the appearance and propagation of jumps in the hydrodynamic fields. This is a familiar problem for nonlinear hydrodynamic equations with piecewise-continuous data, and determining how the relevant weak solutions should be constructed and selected in the present forward--backward saddle-point problem is an interesting mathematical question in its own right. At the same time, the general equations are amenable in principle to numerical treatment, and developing stable numerical schemes would allow the framework to be explored in models where exact solutions are unavailable. Further structural directions include the systematic treatment of symmetry-fixed charge sectors, extensions from dimer states to more general integrable matrix product states, and placing the Quench Action initial measure and the subsystem FCS relation on firmer grounds. More broadly, it would be particularly interesting to determine whether the construction applies to entanglement entropies, and whether their evolution within this framework can be connected to the quasiparticle picture of entanglement spreading. The present results demonstrate that the combination of the Quench Action and BMFT provides a versatile framework for following not only the average hydrodynamic evolution after an integrable quench, but also a broad range of its rare, large-scale fluctuations.

\begin{acknowledgments}

We thank discussions with  K. Chalas,  F. H. L. Essler, F. Hübner, K. Klobas, P. Ruggiero, and R. Travaglino. DXH is particularly grateful to A. Zukaite; and to B. Doyon for his feedback and for inspiring discussions. The work of BB and DXH has been supported by the Royal Society through the University Research Fellowship No.\ 201101.
\end{acknowledgments}

\bibliography{bibliography}  

\appendix

\section{Extensivity of initial fluctuations}

\label{app:overlap}

%======================================================================

The construction of the initial measure in
Sec.~\ref{sec:initial_measure} presupposes that the conserved quantity
under consideration has a nontrivial extensive FCS in the initial state.
The purpose of this appendix is to explain why this is expected to hold
for generic parity-even conserved charges in integrable initial states,
and to characterise the exceptional non-fluctuating cases.

We do not attempt a complete proof here.  A detailed mathematical
treatment of boundary splitting and FCS extensivity for dimer product
states, together with an investigation of extensions to suitable classes
of matrix product states, will be presented in a separate publication.
Here we state the relevant structural consequences and explain how they
apply to integrable quenches.

%---------------------------------------------------------------------

\subsection{Charge symmetry and the fluctuation dichotomy}

%---------------------------------------------------------------------

Let \(Q_L\) be a Hermitian conserved charge on a periodic system of
length \(L\), obtained by summing translates of a fixed finite-range
density.  We say that an initial state is \(Q\)-symmetric if

\begin{equation}
e^{isQ_L}|\Omega_L\rangle
=e^{isq_L}|\Omega_L\rangle
\qquad
\text{for every }s\in\mathbb R,
\label{eq:Q-symmetric-state}
\end{equation}
or, equivalently,
\begin{equation}
Q_L|\Omega_L\rangle=q_L|\Omega_L\rangle,,
\label{eq:Q-eigenstate-condition}
\end{equation}
and otherwise we call it \(Q\)-asymmetric.
For a subsystem \(A_\ell\) of length \(\ell\), define the centred
thermodynamic FCS by
\begin{equation}
f_{0,Q}^{\rm c}(\lambda)
=\lim_{\ell\to\infty}
\frac{1}{\ell}
\ln
\left\langle\Omega\left|
\exp\!\left[
\lambda
\left(
Q_{A_\ell}
-
\langle Q_{A_\ell}\rangle
\right)
\right]
\right|\Omega\right\rangle,
\label{eq:appendix-centered-FCS}
\end{equation}
whenever the limit exists.  More general large-deviation results for
local observables in \(C^*\)-finitely correlated quantum spin-chain
states were established in Ref.~\cite{Ogata}; after blocking, a
dimer product state is a particularly simple member of this class.

For periodic dimer product states and finite-range densities, one finds
the following dichotomy.  Either the state is \(Q\)-symmetric, in which
case
\begin{equation}
\operatorname{Var}_{\Omega_L}(Q_L)=0,
\qquad
f_{0,Q}^{\rm c}(\lambda)=0,
\label{eq:symmetric-zero-FCS}
\end{equation}
or it is \(Q\)-asymmetric and
\begin{equation}
\operatorname{Var}_{\Omega_L}(Q_L)
=L\sigma_Q^2+O(1),
\qquad
\sigma_Q^2>0.
\label{eq:asymmetric-extensive-variance}
\end{equation}
In the latter case the thermodynamic FCS exists and its centred part is
strictly positive,
\begin{equation}
f_{0,Q}^{\rm c}(\lambda)>0
\qquad
(\lambda\neq0).
\label{eq:asymmetric-positive-FCS}
\end{equation}

The basic reason is that correlations in a dimer product state have
strictly finite range after blocking the dimers into single cells.
The variance can then be decomposed into a sum of non-negative
contributions associated with local excitation patterns.  A vanishing
variance density forces every such contribution to vanish and reduces
the action of the centred density to a telescoping, state-relative
coboundary.  On a periodic chain the coboundary contribution telescopes
exactly.  In this case,
\begin{equation}
\bigl(Q_L-\langle Q_L\rangle_{\Omega_L}\bigr)
|\Omega_L\rangle=0,
\label{eq:centered-charge-annihilation}
\end{equation}
i.e.,
\(|\Omega_L\rangle\) is an exact eigenstate of \(Q_L\), with eigenvalue
\(\langle Q_L\rangle_{\Omega_L}\).

If the coboundary condition is not satisfied, at least one of the
orthogonal local-excitation components remains nonzero.  Its translates
occur with a number proportional to the length of the chain.  After
components producing the same excitation pattern have been combined,
different translated patterns are orthogonal in the dimer product state.
Taking the squared norm therefore gives
\begin{equation}
\operatorname{Var}_{\Omega_L}(Q_L)
=\left|
\bigl(Q_L-\langle Q_L\rangle_{\Omega_L}\bigr)
|\Omega_L\rangle
\right|^2
=L\sigma_Q^2+O(1),
\qquad
\sigma_Q^2>0.
\label{eq:variance-excitation-norm}
\end{equation}
Thus the alternative to the exact-eigenstate branch is not merely a
positive variance: the variance is extensive, with a strictly positive
coefficient.  For compatible periodic lengths the relation is typically
exactly proportional to \(L\); the \(O(1)\) term allows harmless finite-size
and blocking conventions.  This extensive fluctuation scaling is also
consistent with quantum central-limit results for mixing finitely
correlated states \cite{Matsui}, although the sharper implication from
\(Q\)-asymmetry to \(\sigma_Q^2>0\) here relies on the exact dimer-product
structure.

The extensive variance suggests a nontrivial extensive centred FCS, but
some additional use of locality is required to make this implication
rigorous.  A linked-cluster calculation provides a size-independent
neighbourhood of analyticity and allows the thermodynamic limit to be
interchanged with differentiation.  Closely related convergent
linked-cluster expansions for the logarithm of Loschmidt amplitudes of
product states under bounded-degree local Hamiltonians were established
in Ref.~\cite{Wild_2023}.  Adapting this construction to imaginary
counting fields gives
\begin{equation}
\bigl(f_{0,Q}^{\rm c}\bigr)''(0)
=\lim_{L\to\infty}\frac{1}{L}
\operatorname{Var}_{\Omega_L}(Q_L)
=\sigma_Q^2>0.
\label{eq:appendix-FCS-curvature}
\end{equation}

Hence \(f_{0,Q}^{\rm c}(\lambda)>0\) near the origin for
\(\lambda\neq0\), and convexity extends this conclusion to every real
nonzero \(\lambda\) for which the thermodynamic FCS exists.

There is also a more elementary route which does not require analyticity.
A fourth-moment estimate produces a finite gain of the moment generating
function at counting fields of order \(L^{-1/2}\).  By itself, this gives
only a subextensive lower bound at fixed counting field.  The gain becomes
extensive once it is combined with the almost-additivity furnished by the
boundary-splitting property described below.

The boundary-splitting property has a further role in relating finite
subsystems to periodic thermodynamic charges.  Removing the finitely many
local densities which cross the boundary of a large interval changes the
logarithm of its moment generating function only by an \(O(1)\) quantity.
Consequently, it does not change the FCS after division by the interval
length.  In particular, the leading extensive FCS of \(Q_{A_\ell}\) agrees
with that obtained by placing the same density on an auxiliary periodic
system of length \(\ell\).  The complex-time locality method underlying
this estimate goes back to Araki \cite{Araki1969}; the finite-volume,
system-size-independent bounds used for the splitting argument follow
from the one-dimensional estimates of
Pérez-García and Pérez-Hernández \cite{Perez-Perez}.
The detailed fourth-moment, linked-cluster and boundary-splitting arguments
are deferred to the separate work mentioned above.

%---------------------------------------------------------------------

%---------------------------------------------------------------------
\subsection{Pair structure and extensive charge fluctuations}

%---------------------------------------------------------------------

We now relate the preceding fluctuation dichotomy to the pair structure of
an integrable initial state.  Let \(|\alpha\rangle\) denote a simultaneous
Bethe eigenstate of the conserved charges and write
\begin{equation}
|\Omega_L\rangle
=\sum_\alpha c_\alpha|\alpha\rangle.
\label{eq:appendix-Bethe-expansion}
\end{equation}

For the standard integrable initial states considered here, nonzero overlaps
are restricted to Bethe states assembled from pairs of quasiparticles with
opposite rapidities, or, more generally, from pairs related by the relevant
spectral involution \cite{piroli2017integrable}.  In models with several
quasiparticle species, the two members of a pair may also belong to
different, for example conjugate, species.  Denote such an allowed pair by
\begin{equation}
p=(a,\vartheta;b,-\vartheta),
\label{eq:appendix-pair-label}
\end{equation}
and define the value of the conserved charge \(Q\) carried by this pair as
\begin{equation}
\mathfrak q_Q(p)
:=
q_a(\vartheta)+q_b(-\vartheta).
\label{eq:appendix-pair-charge}
\end{equation}
If the Bethe state \(\alpha\) contains \(M_\alpha\) pairs
\(p_{\alpha,1},\ldots,p_{\alpha,M_\alpha}\), additivity of the charge gives
\begin{equation}
q_Q(\alpha)
=\sum_{j=1}^{M_\alpha}
\mathfrak q_Q(p_{\alpha,j}).
\label{eq:appendix-multipair-charge}
\end{equation}
Thus the argument includes one-pair, two-pair and all higher multipair
sectors.  Exact factorised overlap formulae provide concrete realisations
of this paired multiparticle structure in several integrable spin chains
\cite{JiangPozsgay2020,GomborPozsgay2021}.

If \(|\Omega_L\rangle\) were \(Q\)-symmetric, all Bethe states having
nonzero overlap with it would necessarily have the same \(Q\)-eigenvalue.
Consequently, the state is \(Q\)-asymmetric whenever its overlap support
contains two states \(\alpha\) and \(\beta\) such that
\begin{equation}
\sum_{j=1}^{M_\alpha}
\mathfrak q_Q(p_{\alpha,j})
\neq
\sum_{j=1}^{M_\beta}
\mathfrak q_Q(p_{\beta,j}).
\label{eq:appendix-charge-separation}
\end{equation}

The pair structure alone does not guarantee this inequality.  In
principle, the nonzero overlaps could be restricted to a special family of
paired configurations on which the total value of \(Q\) is fixed.  We make
the natural, but model-dependent, assumption that no such accidental
restriction occurs, apart from quantum numbers fixed exactly by the
symmetries of the initial state.  In other words, the overlap support is
assumed to contain sufficiently many paired configurations with different
rapidity content.

For a generic parity-even local charge, the one-particle eigenvalue is a
nonconstant function of rapidity.  Its pair contribution
\(\mathfrak q_Q(p)\) therefore changes when the rapidities of the supported
pairs are varied.  Under the preceding assumption, the overlap support then
contains states with different total \(Q\)-eigenvalues.  The initial state
is consequently \(Q\)-asymmetric and, by the fluctuation dichotomy described
above, its variance and centred FCS are extensive.

The expected exceptions are much simpler charges whose values are already
fixed by the quantum numbers of the initial state.  Examples include
magnetisation or particle number in a fixed-charge dimer state, and
topological charge when the overlap support contains only neutral
particle--antiparticle pairs.  Their one-particle or one-pair eigenvalues
are typically independent of rapidity, so varying the rapidity content does
not change the total charge.  Reflection-odd charges constitute a separate
exception: they vanish pairwise on the standard paired support and
annihilate an integrable initial state by definition
\cite{piroli2017integrable}.  Exceptional degeneracies or an unusually
restricted overlap support may produce further model-dependent exceptions.

We therefore expect non-extensive fluctuations to be exceptional among
finite-range parity-even conserved charges in dimer-type integrable initial
states.  They arise naturally when the initial state is an exact eigenstate
of a simple symmetry charge, or when additional degeneracies or restrictions
of the overlap support prevent the charge from separating the supported
Bethe configurations.  Subject to the genericity assumption on the overlap
support described above, genuinely rapidity-dependent parity-even charges
instead belong to the extensive, \(Q\)-asymmetric branch.

The boundary-integrability framework also admits integrable matrix product
states beyond two-site product states \cite{PozsgayPiroliVernier2019}.
Extending the fluctuation dichotomy to suitable members of this larger class
requires additional control of their correlations and is not assumed here.

The detailed boundary-splitting and fluctuation arguments underlying this
conclusion will be presented in a separate publication.  The present
appendix is intended only to clarify the expected range of charges for which
a nontrivial extensive initial FCS, and hence the initial fluctuation measure
used in the main text, can be applicable.
%---------------------------------------------------------------------

\section{Small $\zeta$ regime for Rule 54}
\label{AppSmallT}
The main goal of this appendix is to first explicitly write down \eqref{SmallTFinalSelfConsEqs} and then to show that it is equivalent to the cubic equation \eqref{eq:simplecubic} and demonstrate the uniqueness of the solution. In order to write down the main equation, we first need to define certain shorthands.

\subsection{Logarithmic primitives (canonical forms)}
Define
\[
L(z):=\log\!\Bigl(\frac{1}{1+e^{z}}\Bigr)=-\log(1+e^{z})\,,
\]
which satisfy, for all real $z$,
\begin{equation}
\label{eq:Lsym}
L(-z)=L(z)+z.
\end{equation}
To write down \eqref{SmallTFinalSelfConsEqs}, we use the following objects:
\begin{align}
\ell_0 &:= \log\!\Bigl(\tfrac{1}{1+e^{\mu}}\Bigr)=L(\mu),
&\ell_1 &:= \log\!\Bigl(\tfrac{e^{\mu}}{1+e^{\mu}}\Bigr),\\
\ell_{\chi} &:= \log\!\Bigl(\tfrac{e^{\mu}}{e^{\chi}+e^{\mu}}\Bigr)=\log\!\Bigl(\tfrac{1}{1+e^{\chi-\mu}}\Bigr)=L(\chi-\mu),
&\ell_{2\lambda} &:= \log\!\Bigl(\tfrac{e^{\mu}}{e^{2\lambda}+e^{\mu}}\Bigr)=\log\!\Bigl(\tfrac{1}{1+e^{2\lambda-\mu}}\Bigr)=L(2\lambda-\mu),\\
\tau_{\chi} &:= \log\!\Bigl(\tfrac{1}{1+e^{-\chi+\mu}}\Bigr)=L(\mu-\chi),
&\tau_{2\lambda} &:= \log\!\Bigl(\tfrac{1}{1+e^{-2\lambda+\mu}}\Bigr)=L(\mu-2\lambda).
\label{Logs}
\end{align}
These quantities are not independent; using \eqref{eq:Lsym}, we have
\begin{align}
\label{eq:ell1rel}
\ell_1&=\log\!\Bigl(\tfrac{e^{\mu}}{1+e^{\mu}}\Bigr)=\log\!\Bigl(\tfrac{1}{1+e^{-\mu}}\Bigr)=L(-\mu)=L(\mu)+\mu=\ell_0+\mu,\\
\label{eq:tauchirel}
\tau_{\chi}&=L(\mu-\chi)=L(\chi-\mu)+(\chi-\mu)=\ell_{\chi}+(\chi-\mu),\\
\label{eq:tau2lamrel}
\tau_{2\lambda}&=L(\mu-2\lambda)=L(2\lambda-\mu)+(2\lambda-\mu)=\ell_{2\lambda}+(2\lambda-\mu).
\end{align}

\subsection{Eq. \eqref{SmallTFinalSelfConsEqs} in block form}
Eq. \eqref{SmallTFinalSelfConsEqs} can be written rational expression in $\chi$ involving many repetitions of the same logarithmic combinations and a large square root. The following blocks isolate those repeated combinations.
We define the ``denominator core'' as
\begin{equation}
\label{eq:DenCoreRaw}
\mathrm{DenCore}(\chi):=\ell_0-3\ell_1+6\ell_{\chi}-3\ell_{2\lambda}-2\tau_{\chi}+\tau_{2\lambda}\,.
\end{equation}
Then we define ``product blocks'' as
\begin{align}
\label{eq:Adef}
A(\chi)&:=\mu+\ell_0-2\ell_1+\ell_{\chi},\\
\label{eq:Bdef}
B(\chi)&:=\ell_0-2\ell_1+3\ell_{\chi}-\ell_{2\lambda}-\tau_{\chi},\\
\label{eq:Cdef}
C(\chi)&:=\mu+\ell_1-2\ell_{\chi}+\tau_{\chi},\\
\label{eq:Edef}
D(\chi)&:= -3\ell_{\chi}+3\ell_{2\lambda}+\tau_{\chi}-\tau_{2\lambda},\\
\label{eq:Ddef}
E(\chi)&:=\ell_1-3\ell_{\chi}+2\ell_{2\lambda}+\tau_{\chi}-\tau_{2\lambda},\\
F(\chi)&:=\mu-2\ell_{\chi}+\ell_{2\lambda}+\tau_{\chi},\\
G(\chi)&:=\mu+\ell_{\chi}-2\ell_{2\lambda}+\tau_{2\lambda}.
\end{align}
and certain product expressions of the above quantities as
\begin{equation}
\label{eq:Sdef}
S(\chi):=A(\chi)B(\chi)-C(\chi)E(\chi)\,, \quad \text{and}\quad U(\chi):=C(\chi)F(\chi)-A(\chi)G(\chi)\,,
\end{equation}
and $\mathcal{Q}$ as
\begin{equation}
\label{eq:Qstructured}
\mathcal{Q}(\chi)=U(\chi)^2+4\lambda\,D(\chi)S(\chi)+4\lambda^2D(\chi)^2.
\end{equation}
The last bit we need is $\mathcal{N}$, which is \begin{equation}
\label{eq:Ncompact}
\quad \mathcal{N}(\chi)=-S(\chi)-2\lambda\,\mathrm{DenCore}(\chi).\quad
\end{equation}
Using these objects, it is possible to express \eqref{SmallTFinalSelfConsEqs} 
via 
\begin{equation}
\label{eq:Eq1NewCompact}
\chi
= -\frac{\mathcal{N}(\chi)+\sqrt{\mathcal{Q}(\chi)}}{2\,\mathrm{DenCore}(\chi)}\,.
\end{equation}

\subsection{Showing the reduction to the cubic equation}
Consider first the rewriting of the cubic equation \eqref{eq:simplecubic}
\begin{equation}
\label{eq:simple}
\begin{split}
&-\mu-\ln\!\Bigl(\tfrac1{1+e^{\mu}}\Bigr)+2\ln\!\Bigl(1-\tfrac1{1+e^{\mu}}\Bigr)-\ln\!\Bigl(1-\tfrac1{1+e^{(\mu-\chi)}}\Bigr)\\
&=\lambda-\mu-\ln\!\Bigl(\tfrac1{1+e^{(\mu-\chi)}}\Bigr)+2\ln\!\Bigl(1-\tfrac1{1+e^{(\mu-\chi)}}\Bigr)-\ln\!\Bigl(1-\tfrac1{1+e^{(\mu-2\lambda)}}\Bigr)\,.
\end{split}
\end{equation}
Using the identities
\[
1-\frac{1}{1+e^{a}}=\frac{e^{a}}{1+e^{a}},\qquad
\log\!\Bigl(\frac{e^{a}}{1+e^{a}}\Bigr)=a-\log(1+e^{a}),\qquad
\log\!\Bigl(\frac1{1+e^{a}}\Bigr)=-\log(1+e^{a})\,,
\]
it is immediate to see that \eqref{eq:simple} is equivalent to
\begin{equation}
\label{eq:res0}
\mu-\ln(1+e^{\mu})+2\ln(1+e^{\chi-\mu})
=\lambda-\chi+\ln(1+e^{2\lambda-\mu}).
\end{equation}
Now, defining the "residual", as
\begin{equation}
\label{eq:Rdef}
R(\chi):=\mu-\log(1+e^{\mu})+2\log(1+e^{\chi-\mu})-\Bigl(\lambda-\chi+\log(1+e^{2\lambda-\mu})\Bigr)\,,
\end{equation}
the cubic equations \eqref{eq:simple} is equivalent to
\begin{equation}
\label{eq:Rzero}
R(\chi)=0.
\end{equation}

In the following we show that various blocks of \eqref{eq:Eq1NewCompact} have specific structures.
As a next step, we first
reconsider \eqref{eq:DenCoreRaw}, and substitute \eqref{eq:ell1rel}--\eqref{eq:tau2lamrel} into it, yielding
\[
\mathrm{DenCore}(\chi)
=\ell_0-3(\ell_0+\mu)+6\ell_{\chi}-3\ell_{2\lambda}-2(\ell_{\chi}+\chi-\mu)+(\ell_{2\lambda}+2\lambda-\mu)\,,
\]
and simplifying to
\begin{equation}
\label{eq:DenCoreSimplified}
\mathrm{DenCore}(\chi)=-2\ell_0+4\ell_{\chi}-2\ell_{2\lambda}-2\chi+2\lambda.
\end{equation}
Now rewriting $\ell_0,\ell_{\chi},\ell_{2\lambda}$ by their original definitions
\[
\ell_0=-\log(1+e^{\mu}),\quad
\ell_{\chi}=-\log(1+e^{\chi-\mu}),\quad
\ell_{2\lambda}=-\log(1+e^{2\lambda-\mu}),
\]
upon substituting into \eqref{eq:DenCoreSimplified}, we obtain the crucial link between the residual (cubic equation) and the denominator of \eqref{eq:Eq1NewCompact}:
\begin{equation}
\label{eq:DenCoreIs2R}
\quad \mathrm{DenCore}(\chi)=2R(\chi). \quad
\end{equation}

Because $\mathrm{DenCore}(\chi)=2R(\chi)$, any $\chi^\ast$ that solves the simplified residual equation $R(\chi)=0$ necessarily satisfies
\[
\mathrm{DenCore}(\chi^\ast)=0.
\]
In particular, the original fixed-point formula \eqref{eq:Eq1NewCompact} is written as a quotient, and anticipating that it has the same solution as the cubic equation, it cannot be evaluated at $\chi^\ast$ in the strict sense: it becomes a $0/0$-type expression. Therefore we shall interpret the solutions of \eqref{eq:Eq1NewCompact}  via its \emph{cleared form} (equivalently, the continuous extension after canceling common factors). That is, we regard $\chi$ as an \emph{extended solution} if it satisfies
\begin{equation}
\label{eq:Eq1NewCleared}
2\chi\,\mathrm{DenCore}(\chi)+\mathcal N(\chi)+\sqrt{\mathcal Q(\chi)}=0,
\end{equation}
where $\sqrt{\cdot}$ is the principal real square root.
This equation makes sense even when $\mathrm{DenCore}(\chi)=0$ (provided $\mathcal Q(\chi)\ge 0$), and it is algebraically equivalent to \eqref{SmallTFinalSelfConsEqs} whenever \eqref{SmallTFinalSelfConsEqs} is strictly well-defined (i.e.\ whenever $\mathrm{DenCore}(\chi)\neq 0$).

\paragraph{Reduction on the removable-singularity (cubic) branch.}
On the cubic branch we impose $\mathrm{DenCore}(\chi)=0$, i.e. $R(\chi)=0$ by \eqref{eq:DenCoreIs2R}. On this branch the actual radicand \eqref{eq:Qstructured} simplifies further: one checks (algebraically) that
\begin{equation}
\label{eq:Qfactor_on_cubic}
\mathcal Q(\chi)-\bigl(S(\chi)+2\lambda D(\chi)\bigr)^2
=4\chi\lambda\,R(\chi)\,G(\chi;\mu),
\end{equation}
with
\[G(\chi;\mu):=\chi-2\log(1+e^{\mu})+2\log(e^{\chi}+e^{\mu}).\]
Hence, on $R(\chi)=0$, one has $\mathcal Q(\chi)=\bigl(S(\chi)+2\lambda D(\chi)\bigr)^2$ and therefore
\begin{equation}
\label{eq:sqrtQ_on_cubic}
\sqrt{\mathcal Q(\chi)}=\bigl|S(\chi)+2\lambda D(\chi)\bigr|\qquad\text{on the branch }R(\chi)=0.
\end{equation}
Substituting $\mathrm{DenCore}(\chi)=0$ into \eqref{eq:Eq1NewCleared} gives $-S(\chi)+\sqrt{\mathcal Q(\chi)}=0$, so \eqref{eq:sqrtQ_on_cubic} implies $-S(\chi)+|S(\chi)+2\lambda D(\chi)|=0$ on the cubic branch. In particular, the removable-singularity solution is selected by $R(\chi)=0$ and is therefore governed by the cubic equation \eqref{eq:simplecubic}. This cubic has a unique real solution as shown below.

\subsection{Properties of the cubic equation and existence and uniqueness of \eqref{eq:Eq1NewCompact}}
From $R(\chi)=0$ one can rewrite the  complicated equation \eqref{SmallTFinalSelfConsEqs} into a compact exponential form.
Introducing
\[
y:=e^{\chi-\mu}>0\,,
\]
a direct simplification of \eqref{eq:simple} yields
\[
\log y+2\log(1+y)=\log\Bigl(e^{\lambda-2\mu}(1+e^{2\lambda-\mu})(1+e^{\mu})\Bigr),
\]
and exponentiating it gives
\begin{equation}
\label{eq:yEqK}
\quad y(1+y)^2=K,\qquad K:=e^{\lambda-2\mu}(1+e^{2\lambda-\mu})(1+e^{\mu}).
\end{equation}
Expanding $y(1+y)^2=y+2y^2+y^3$ yields the cubic polynomial
\begin{equation}
\label{eq:CubicInY}
\quad y^3+2y^2+y-K=0.
\end{equation}
Equivalently, in $u=e^{\chi}$ (so $y=ue^{-\mu}$) this is
\[
 u^3+2e^{\mu}u^2+e^{2\mu}u-e^{\lambda+\mu}(1+e^{2\lambda-\mu})(1+e^{\mu})=0.
\]

We can easily demonstrate the uniqueness of the cubic solutions as well.
Since $K>0$ for all real $\mu,\lambda$, one can define $g(y):=y(1+y)^2$ on $(0,\infty)$ such that
\[
g'(y)=1+4y+3y^2>0,
\]
i.e., $g$ is strictly increasing with $g(0^+)=0$ and $g(y)\to\infty$ as $y\to\infty$. Consequently, \eqref{eq:yEqK} has a unique solution $y^*>0$, and therefore \eqref{eq:CubicInY} has a unique positive root. This yields a unique real
\[
\chi^*:=\mu+\log y^*
\]
solving $R(\chi)=0$, and therefore a unique extended solution on the removable-singularity/cubic branch.

\subsection{Excluding additional solutions}

As said, the solution of \eqref{SmallTFinalSelfConsEqs} is not unique. However, it is possible to show that the only physical solution is unique, and is given by the cubic equation. In particular, we study the values of the proportionality constants (for simplicity) and we find that additional solutions always result either in complex,  or in divergent or zero values for these constants. Complex proportionality constants imply complex values for the $H$ fields and this fact allows us the get rid of these solutions, as we can naturally impose real-valuedness for the auxiliary hydrodynamic fields with real terminal conditions.
We also exclude cases when these constants are infinite (forcing the $H$ field to be zero in a region) and zero (meaning the absence of jumps for the $H$ field across an interface. These scenarios were argued not to be consistent with the terminal condition for $H$.

%Concretely, focusing on the explicit expression for the constants $\alpha$, and $\gamma$, for real $(\mu,\lambda,\chi)$ all logarithms entering these expressions are real, so the only source of complex values is the square-root discriminant that appears (with opposite sign) in $\alpha$ and $\gamma$.
%We therefore call a strict fixed point $\chi$ to be physical if $\alpha(\mu,\lambda,\chi)$ and $\gamma(\mu,\lambda,\chi)$ are both real (equivalently, the shared discriminant is nonnegative and the denominators are nonzero).

The expressions for $\alpha$ and $\gamma$ can be written as follows by using the log primitives
$\ell_0,\ell_1,\ell_\chi,\ell_{2\lambda},\tau_\chi,\tau_{2\lambda}$ and by the linear factor
\[
P(\chi):=\ell_0-3\ell_1+3\ell_\chi-\tau_\chi,\qquad
\]
We also define furthermore the non-root numerator
\begin{align*}
M(\chi):={}&
\mu\,\ell_0-3\mu\,\ell_1-6\lambda\,\ell_\chi+6\mu\,\ell_\chi+\ell_0\ell_\chi-3\ell_\chi^2
+6\lambda\,\ell_{2\lambda}-3\mu\,\ell_{2\lambda}-2\ell_0\ell_{2\lambda}+3\ell_1\ell_{2\lambda}\\
&\quad
+2\lambda\,\tau_\chi-2\mu\,\tau_\chi-\ell_1\tau_\chi+4\ell_\chi\tau_\chi-\ell_{2\lambda}\tau_\chi-\tau_\chi^2
-2\lambda\,\tau_{2\lambda}+\mu\,\tau_{2\lambda}+\ell_0\tau_{2\lambda}-2\ell_1\tau_{2\lambda}+\ell_\chi\tau_{2\lambda}.
\end{align*}
With these abbreviations, one can write
\[
\alpha(\mu,\lambda,\chi)=\frac{M(\chi)-\tfrac12\sqrt{\Delta_{\alpha\gamma}(\chi)}}{P(\chi)\,Q},
\qquad
\gamma(\mu,\lambda,\chi)=\frac{-M(\chi)+\tfrac12\sqrt{\Delta_{\alpha\gamma}(\chi)}}{Q\,(-D(\chi))},
\]
where the shared discriminant is
\[
\Delta_{\alpha\gamma}(\chi)=16\lambda\,P(\chi)\,Q\,D(\chi)+4\bigl(U(\chi)-2\lambda\,(F(\chi)-G(\chi))\bigr)^2.
\]
In particular, for real $(\mu,\lambda,\chi)$ the pair $(\alpha,\gamma)$ is real-valued if and only if $\Delta_{\alpha\gamma}(\chi)\ge 0$ and the denominators are nonzero.

Using a multi-start root finder (wrt. the initial value for the root-finding algorithm) to enumerate strict real fixed points of \eqref{SmallTFinalSelfConsEqs} and then filtering by real-valuedness and the non-zero finiteness of $\alpha$ and $\gamma$, we performed a uniform parameter scan on the grid
\[
\lambda\in[-5,5],\qquad \mu\in[-5,5],\qquad \Delta\lambda=\Delta\mu=10^{-2}.
\]
On this grid we did not encounter any parameter point for which there exists a strict real fixed point with nonreal, finite and non-zero $(\alpha,\gamma)$ that would pass the physical filter; moreover, whenever additional strict real fixed points occur, their associated $(\alpha,\gamma)$ become complex and are thus excluded. To make such a claim, we needed to impose certain numerical tolerance criterion. To be non zero and finite, we demanded that
\beq
10^{-5}< |\alpha|,|\gamma| < 10^4\,,
\eeq
and we declared these objects to be real if,
\beq
|\text{Im}(\alpha)|,|\text{Im}(\gamma)| <10^{-24}\,.
\eeq
The numerical finding  provides strong numerical evidence that the cubic/removable-singularity solution is the unique physical solution, even though \eqref{SmallTFinalSelfConsEqs} may admit additional strict real fixed points for some $(\mu,\lambda)$.

\section{Large $\zeta$ regime for Rule 54}
\label{AppLargeT}

We first present the equations that define the solutions of the unknowns $\chi$ and $\upsilon$.
To write the equations in a compact way, we
recall the function
$$
L(z) := \log\!\Bigl(\frac{1}{1+e^{z}}\Bigr) = -\log(1+e^{z})\,,
$$
and the identity
$$
L(-z) = L(z) + z
$$
for all real $z$. We recall from \eqref{Logs} and introduce further abbreviations such as
$$
\tau_\upsilon := L(\upsilon-\mu),\qquad \tau_\chi := L(\chi-\mu),\qquad \ell_1 := L(-\mu).
$$
and then we can eventually write the equations \eqref{LargeTFinalSelfConsEqs} into the compact form

\beq
\chi = \frac{\chi\,\lambda\,N_1}{D_1}\,,\quad
\upsilon = \frac{2\lambda\,N_2}{D_2}\,,
\label{LargeTFinalSelfConsEqsApp}
\eeq
with
$$
N_1 = \chi-\upsilon + 2(\tau_\upsilon-\tau_\chi),
$$
$$
D_1 = \chi^2 + \chi\bigl(-\upsilon + \tau_\upsilon-2\tau_\chi+\ell_1\bigr) + \upsilon\bigl(\tau_\chi-\ell_1\bigr),
$$
and with
$$
N_2 = \chi(\ell_1-\tau_\upsilon) + \upsilon(\tau_\chi-\ell_1),
$$
$$
D_2 = \chi\bigl(\upsilon-\chi + 2\tau_\chi-\tau_\upsilon-\ell_1\bigr) + \upsilon(\ell_1-\tau_\chi).
$$
Now, a direct comparison shows
$$
D_2 = -\Bigl(\chi^2 + \chi\bigl(-\upsilon + \tau_\upsilon-2\tau_\chi+\ell_1\bigr) + \upsilon\bigl(\tau_\chi-\ell_1\bigr)\Bigr) = -D_1\,,
$$
hence
$$
\upsilon = -\frac{2\lambda\,N_2}{D_1}.
$$
Assuming that the expressions \eqref{LargeTFinalSelfConsEqsApp} are well-defined and that $\chi\neq 0$,  from
$$
\chi = \frac{\chi\,\lambda\,N_1}{D_1}
$$
we obtain
$$
1 = \frac{\lambda N_1}{D_1},\qquad\text{hence}\qquad \lambda = \frac{D_1}{N_1}.
$$
In particular, this step requires $N_1\neq 0$.
Substituting $$\lambda=D_1/N_1$$ into the second expression in \eqref{LargeTFinalSelfConsEqsApp}
yields
$$
\upsilon = -\frac{2N_2}{N_1}\,,\quad \text{or}\quad 
\upsilon N_1 + 2N_2 = 0.
$$
This gives
$$
0 = \upsilon\bigl(\chi-\upsilon+2(\tau_\upsilon-\tau_\chi)\bigr) + 2\bigl(\chi(\ell_1-\tau_\upsilon) + \upsilon(\tau_\chi-\ell_1)\bigr)
= (\chi-\upsilon)\bigl(\upsilon + 2(\ell_1-\tau_\upsilon)\bigr)\,,
$$
thus any well-defined solution with $\chi\neq 0$ must satisfy
$$
(\chi-\upsilon)\bigl(\upsilon + 2(\ell_1-\tau_\upsilon)\bigr)=0.
$$

It is easy to exclude the possibility of $\chi=\upsilon$. Namely, if $\chi=\upsilon$, then $\tau_\upsilon=\tau_\chi$ and therefore
$$
N_1 =\chi-\upsilon + 2(\tau_\upsilon-\tau_\chi) = 0.
$$
However, the first equation of \eqref{LargeTFinalSelfConsEqs} in the reduced form
$$
1=\frac{\lambda N_1}{D_1}
$$
cannot hold with $N_1=0$ when the right-hand side is well-defined.
Hence, among well-defined solutions with $\chi\neq 0$, the option $\chi=\upsilon$ is  not possible.

The only remaining possibility is therefore
$$
\upsilon + 2(\ell_1-\tau_\upsilon)=0,
$$
that is,
$$
\upsilon + 2\Bigl(L(-\mu)-L(\upsilon-\mu)\Bigr)=0.
$$
To proceed, we first define
$$
J(\upsilon) := \upsilon + 2\Bigl(L(-\mu)-L(\upsilon-\mu)\Bigr).
$$
and consequently $J(0)=0$.
Differentiating $J$ using that
$$
L'(z) = -\frac{e^z}{1+e^z} \in (-1,0)
$$
(for all real $z$), we get
$$
J'(f) = 1 - 2L'(\upsilon-\mu) \in (1,3),
$$
therefore $J$ is strictly increasing on $\mathbb{R}$ and has at most one root. Since $H(0)=0$, the unique root is
$$
\upsilon=0.
$$

Now substituting $\upsilon=0$ into \eqref{LargeTFinalSelfConsEqs}, one finds that the denominator satisfies
$$
D_1 = \chi N_1
$$
for $\upsilon=0$ (since in this case $\tau_\upsilon=\ell_1$). Therefore \eqref{LargeTFinalSelfConsEqs} becomes
$$
\chi = \frac{\chi\lambda N_1}{\chi N_1} = \lambda,
$$
again provided the expression is well-defined. 
To summarise, assuming nonzero denominators in \eqref{LargeTFinalSelfConsEqs},   the system has the unique real solution
$$
 \chi=\lambda\,, \quad \upsilon=0\,.
$$

\section{Proof for the overlapping branches of solution in Rule 54}
\label{ProofForBranches}

We prove that the two branches in \eqref{R54SolWithBranches} always overlap for any real $\mu$ and $\lambda$, which make it possible to argue for a unique and continuous physical solution \eqref{R54SolWithoutBranches}. We need to show that 
\beq
\begin{split}
\text{i)}&\qquad(v^\text{eff}_{\chi^*})^{-1} \geq (v^\text{eff}_\lambda)^{-1} \Leftrightarrow n_1 \geq n_\lambda\\
\text{ii)}&\qquad f_\text{GGE}(\lambda,\mu)-\frac{1}{2}f_\text{GGE}(2\lambda,\mu) \geq 2\frac{\textgoth{m}}{v^\text{eff}_{\chi^*}}
\end{split}
\eeq
both hold simultaneously for any real $\mu$ and $\lambda$.
Starting with condition i), we recall that 
\beq
n_1=\frac{1}{1+e^{\mu-\chi}}\,,\quad \text{and}\quad n_\lambda=\frac{1}{1+e^{\mu-\lambda}}\,,
\eeq
and we begin by defining
\beq
V(x;\mu):=(x-\mu)+2\log\!\bigl(1+e^{x-\mu}\bigr),\qquad
\log K(\lambda,\mu):=\lambda-2\mu+\log(1+e^\mu)+\log\!\bigl(1+e^{2\lambda-\mu}\bigr),
\label{VAndLogK}
\eeq
and
\[
u(x;\lambda,\mu):=V(x;\mu)-\log K(\lambda,\mu).
\]
Let $\chi$ denote the unique real root of $u(x;\lambda,\mu)=0$ (equivalently, the unique solution of the cubic in $y=e^{x-\mu}$). We now show that for $\lambda >0$
\[
\lambda<\chi<2\lambda\,,
\]
from which i) follows for positive $\lambda$.

We differentiate first
\[
u'(x)=\frac{\partial}{\partial x}V(x;\mu)=1+\frac{2}{1+e^{-(x-\mu)}}\in(1,3),
\]
yielding that $u(x)$ is strictly increasing and has at most one real root.
Next, we evaluate $u$ at $x=\lambda$ and $x=2\lambda$.
At the former point,
\[
u(\lambda)=\log\!\Big(\frac{e^{\mu}\bigl(1+e^{\lambda-\mu}\bigr)^2}{(1+e^\mu)(e^{2\lambda}+e^\mu)}\Big)\,,
\]
and since
\[
(1+e^\mu)(e^{2\lambda}+e^\mu)-e^\mu(1+e^{\lambda-\mu})^2=(e^\lambda-1)^2>0,
\]
the fraction inside the logarithm is $<1$, hence $u(\lambda)<0$.
When $x=2\lambda$, we have
\[
u(2\lambda)=\lambda+\log\!\Big(\frac{e^{2\lambda}+e^\mu}{1+e^\mu}\Big).
\]
Because $e^{2\lambda}>1$, we have that $(e^{2\lambda}+e^\mu)/(1+e^\mu)>1$, therefore the logarithm is $>0$ and hence $u(2\lambda)>\lambda>0$.
By strict monotonicity, $u(\lambda)<0<u(2\lambda)$ implies that the unique root $\chi$ satisfies $\lambda<\chi<2\lambda$.

For $\mu\in\mathbb{R}$ and $\lambda<0$, shall instead show that
\[
\lambda<x<0.
\]
from which i) again holds, now for $\lambda<0$.
As before, we take $u(x)$ (strictly increasing as a function of $x$) and now evaluate it at the points $\lambda$ and $0$.

Similarly to the case of $\lambda>0$, also for negative values we have that $u(\lambda)<0$.
Then, at $x=0$ one may simplify $u(0)$ to
\[
u(0)=\log(1+e^{-\mu})-\lambda-\log\!\bigl(1+e^{2\lambda-\mu}\bigr).
\]
For $\lambda<0$ we have $e^{2\lambda-\mu}<e^{-\mu}$, consequently,
\[
\frac{1+e^{-\mu}}{1+e^{2\lambda-\mu}}>1,
\qquad\text{hence}\qquad
\log\!\Big(\frac{1+e^{-\mu}}{1+e^{2\lambda-\mu}}\Big)>0.
\]
Given that $-\lambda>0$, we eventually find that
\[
u(0)=\log\!\Big(\frac{1+e^{-\mu}}{1+e^{2\lambda-\mu}}\Big)-\lambda>0.
\]
Thus $u(\lambda)<0<u(0)$, and by strict monotonicity the unique root satisfies $\lambda<x<0$.

Now we move to condition ii) 
and first recall that the cubic equation for $\chi$ ($u(x)=0$) can be recast as
\[
V(x;\mu)=\log K(\lambda,\mu).
\]
To prove
\[
 f_{\mathrm{GGE}}(\mu,\lambda)-\tfrac12 f_{\mathrm{GGE}}(\mu,2\lambda)
 \;\ge\; 2\,\frac{\mathfrak{m}}{v^\text{eff}_{\chi^*}}\,,
\]
we first directly rearrange the lhs. using \eqref{GGEFCSForBranches} and \eqref{VAndLogK}, which gives
\[
 f_{\mathrm{GGE}}(\mu,\lambda)-\tfrac12 f_{\mathrm{GGE}}(\mu,2\lambda)
 = \mu+2\log(1+e^{\lambda-\mu})-\log(1+e^{\mu})-\log(1+e^{2\lambda-\mu})
 = V(\lambda;\mu)-\log K(\lambda,\mu).
\]
Since $\chi$ satisfies $V(x;\mu)=\log K(\lambda,\mu)$, we obtain
\[
 f_{\mathrm{GGE}}(\mu,\lambda)-\tfrac12 f_{\mathrm{GGE}}(\mu,2\lambda)=V(\lambda;\mu)-V(\chi;\mu).
\]
Next,
\[
\frac{\text{d}}{\text{d} t}V(t;\mu)=V'(t;\mu)=1+\frac{2}{1+e^{\mu-t}},
\qquad\text{hence}\qquad
V'(\chi;\mu)=1+2n_1.
\]
Using the identity $\mathfrak{m}=\lambda-\chi^*$ \eqref{Slope_Gen} and the definition of the effective velocity $v^\text{eff}_{\chi^*}=2/(1+2n_1)$, we rewrite
\[
2\,\frac{\mathfrak{m}}{v^\text{eff}_{\chi^*}}=2(\lambda-\chi)\frac{1+2n_1}{2}=(\lambda-x)\,V'(\chi;\mu).
\]
Finally, 
\beq
\frac{\text{d}^2}{\text{d} t^2}V(t;\mu)=V''(t;\mu)=2\,\sigma(t-\mu)(1-\sigma(t-\mu))\ge 0
\eeq
with $\sigma(s)=1/(1+e^{-s})$, therefore $V(\cdot;\mu)$ is convex. Consequently,
\[
V(\lambda;\mu)\ge V(\chi;\mu)+V'(\chi;\mu)(\lambda-\chi),
\]
which rearranges to
\[
V(\lambda;\mu)-V(\chi;\mu)\ge (\lambda-x)V'(\chi;\mu)=2\,\frac{\mathfrak{m}}{v^\text{eff}_{\chi^*}}
\]
yielding the desired inequality and proving ii).

\end{document}